\documentclass[aps,pre,twocolumn,reprint,amssymb,amsmath,nobibnotes,nofootinbib,superscriptaddress,floatfix]{revtex4}
\usepackage[dvips]{graphicx}
\usepackage{epsfig}
\usepackage{framed}
\usepackage{color}
\usepackage{xcolor}
\usepackage{textcomp}
\usepackage{setspace}
\usepackage{morefloats}
\usepackage[T1]{fontenc}
\usepackage[utf8]{inputenc}
\usepackage{hyperref}
\usepackage{cleveref}

\begin{document}

\title{Ordering kinetics in $q$-state random-bond clock model: Role of Vortices and Interfaces}

\author{Swarnajit Chatterjee}
\email{sspsc5@iacs.res.in}
\affiliation{School of Mathematical \& Computational Sciences, Indian Association for the Cultivation of Science, Kolkata 700032, India.}

\author{Sabyasachi Sutradhar}
\affiliation{Yale University, 266 Whitney Avenue, New Haven, CT, 06511.}

\author{Sanjay Puri}
\email{puri@mail.jnu.ac.in}
\affiliation{School of Physical Sciences, Jawaharlal Nehru University, New Delhi 110067, India.}

\author{Raja Paul}
\email{raja.paul@iacs.res.in}
\affiliation{School of Mathematical \& Computational Sciences, Indian Association for the Cultivation of Science, Kolkata 700032, India.}

\date{\today}


\begin{abstract}
In this article, we present a Monte Carlo study of phase transition and coarsening dynamics in the non-conserved two-dimensional random-bond $q$-state clock model (RBCM) deriving from a pure clock model [{\bf{Phys. Rev. E 98, 032109 (2018)}}]. Akin to the pure clock model, RBCM also passes through two different phases when quenched from a disordered initial configuration representing at infinite temperature. Our investigation of the equilibrium phase transition affirms that both upper ($T_c^1$) and lower ($T_c^2$) phase transition temperatures decrease with bond randomness strength $\epsilon$. Effect of $\epsilon$ on the non-equilibrium coarsening dynamics is investigated following independent rapid quenches in the quasi-long range ordered (QLRO, $T_c^2 < T < T_c^1$) regime and long-range ordered (LRO, $T<T_c^2$) regime at temperature $T$. We report that the dynamical scaling of the correlation function and structure factor are independent of $\epsilon$ and the presence of quenched disorder slows down domain coarsening. Coarsening dynamics in both LRO and QLRO regimes are further characterized by power-law growth with disorder-dependent exponents within our simulation time scales. The growth exponents in the LRO regime decreases from 0.5 in the pure case to 0.22 in the maximum disordered case, whereas the corresponding change in the QLRO regime happens from 0.45 to 0.38. We further explored the coarsening dynamics in the bond-diluted clock model and in both the models, the effect of the disorder is more significant for the quench in the LRO regime compared to the QLRO regime.
\end{abstract}

\maketitle
\section{Introduction}

In statistical physics, phase transitions exhibited by a large class of model systems are either first-order or second-order types. Apart from these phase transitions manifested in most spin models, a specific type of phase transition called the Berezinskii-Kosterlitz-Thouless (BKT) transition \cite{berezinskii70,thouless73,kosterlitz74} or more generally the Kosterlitz-Thouless (KT) phase transition is observed in various physical systems and can be explained by the two-dimensional XY model. The KT transition does not involve any symmetry breaking and proceeds via the binding and unbinding of topological defects, the vortex-antivortex pairs. According to Mermin-Wagner theorem \cite{mermin}, in continuous systems spin-wave excitation easily destroys any long-range ordering; but Kosterlitz-Thouless showed that a transition indeed takes place at a finite $T$. There exist two phases in the XY model: a low-temperature phase, characterized by the bound vortex-antivortex pairs with quasi-long range order (QLRO), where the spin-spin correlation function decays algebraically and a high-temperature disordered phase characterized by free vortices where the correlation-function decays exponentially.

The $q$-state clock model is a discrete version of the generalized XY model ($q \to \infty$) where the clock spin vectors can draw only specific angles governed by the value of $q$. In $d=2$, this model exhibits a second-order phase transition between a high-temperature paramagnetic phase and a low-temperature long-range ordered (LRO) phase for $2 \leqslant q \leqslant 4$. However, for $q \geqslant 5$, the system displays two transitions at temperatures $T_c^1$ and $T_c^2$ ($T_c^1>T_c^2$) \cite{kadanoff77,elitzur,domany,cardy,tobochnik82,plascak2010,baek2010,swarna2018}, separated by a topological non-trivial KT phase with quasi long-range order (QLRO) emerges between the high-temperature disordered phase and low-temperature LRO phase. There are concluding evidence regarding the nature of the transitions occurring at $T_c^1$ and $T_c^2$ for $q>4$ which confirm that these are indeed Kosterlitz-Thouless (KT) type phase transitions \cite{kadanoff77,surungan2005,tomita2002,plascak2010,miyashita78,landau86,ono91,
okabe2002,tomita2001,rastelli2004,kim2010,Wu2012,li}.

XY model under various kinds of disorders has been studied extensively in literature \cite{surungan2005,leonel2003,alonso2010,harris74,wysin2005,manoj2017} concluding that disorder has strong effects on the KT phase transition. It has been also argued that quenched disorder has substantial effects on the transition temperatures ($T_c^1$ and $T_c^2$) of the $q$-state clock model. For instance, the bond-diluted six-state clock model shows a systematic decrease in the transition temperatures with an increased concentration of missing bonds \cite{surungan2005}. Another study of the random-bond six-state clock model where bond randomness is introduced by drawing the coupling coefficients from a Gaussian distribution shows that the critical temperature of the system gets reduced by the disorder, however, keeping the nature of transition unaltered \cite{Wu2012}. 

A system becomes thermodynamically unstable when quenched below the critical temperature. Due to this quench, the subsequent evolution of the system is characterized by the formation and growth of the domains. The kinetics of phase ordering of a statistical system is the process through which the far-from-equilibrium disordered system tries to attain a spontaneously magnetized equilibrium state by separating into domains. A careful understanding of the process involves investigations of domain morphology, scaling behavior and the asymptotic domain growth law of that system \cite{Bray94,puri-wadhawan}. When systems are cooled through the transition temperatures, interconnected domains of the two equilibrium phases form and coarsen to decrease the total interfacial area and these domains are characterized by a growing characteristic length scale $R(t)$. $R(t)$ typically grows algebraically with time $t$, $R(t)$ $\sim$ $t^n$ where $n$ is typically known as the `growth exponent'. It is well established that a system with non-conserved order parameter obeys Lifshitz-Cahn-Allen (LCA) growth law, $R(t)$ $\sim$ $t^\frac{1}{2}$, whereas for a system with conserved order parameter, Lifshitz-Slyozov (LS) growth law $R(t)$ $\sim$ $t^\frac{1}{3}$ describes the domain growth process \cite{ls,lca,oono-puri,BR94,RB95,puri95}. 

Domain growth in $q$-state clock model with non-conserved order parameter and in the absence of disorder obeys the LCA growth law \cite{swarna2018,Bray94,puri-wadhawan,corberi2006}. Here domain coarsening occurs via the elimination of both domain interfaces and vortices \cite{corberi2006}. A good understanding of domain growth kinetics in pure systems, shifted the focus in recent years toward the domain growth kinetics in disordered systems \cite{manoj2017,puri91,puri93,rp2004,rp2005,henkel2006,rp2007,henkel2008,puri2010,puri2011,puri2012,manoj2014,
manoj17} due to greater experimental relevance. However, establishing the true nature of the ordering kinetics in disorder systems has been debated over the decades. In the random-bond Ising model (RBIM), earlier studies by Paul $et~al$ suggested power-law growth with disorder dependent exponents \cite{rp2004,rp2005}, however, recent studies of coarsening in RBIM and of topological defects in oscillating systems with quenched disorder argued a crossover from a faster power-law growth to a slower logarithmic growth in the asymptotic limit \cite{puri2010,puri2011,puri2012,reichhardt}. The ordering kinetics of the random-bond XY model (RBXYM) \cite{manoj2017} in $d = 2$ shows an algebraic growth with a disorder-dependent exponent, although, in $d = 3$ the asymptotic growth law appears to be logarithmic. We expect a logarithmic growth even for $d=2$ RBXYM but could not observe it within our simulation time scales. A clock model with the disorder is a highly significant classical statistical model as it interpolates between the scalar Ising model and the vector XY model \cite{lupo} - nevertheless, to the best of our knowledge, a domain growth kinetics in disordered clock model is still lacking. 

In this paper, we present a study of the effect of bond randomness on the equilibrium phase transition temperatures and phase ordering kinetics in the $q$-state random-bond clock model (RBCM) with $q$ = 6 and 9 in $d=2$. As described above, for these $q$-values, we have two transitions, one from disordered to QLRO phase at $T_c^1$ and another from QLRO to LRO phase at $T_c^2$. After equilibrating the system via Wolff single cluster algorithm \cite{wolff89}, we quantify $T_c^1$ and $T_c^2$ as a function of $\epsilon$. The ordering kinetics is then studied by rapidly quenching the system in both LRO ($T<T_c^2$) and QLRO ($T_c^2<T<T_c^1$) regimes separately and the evolution is studied via Metropolis algorithm \cite{metropolis53}. The main results of our investigation of the RBCM are summarized as follows:

(a) Both the upper ($T_c^1$) and lower ($T_c^2$) transition temperatures are decreasing functions of the disorder strength $\epsilon$.

(b) Ordering kinetics in RBCM for a temperature quench in the LRO regime ($T<T_c^2$) is characterized by well-defined sharp domain boundaries with domain size shrinking with $\epsilon$. Dynamical scaling is independent of disorder, and therefore universal. Similar to RBXYM \cite{manoj2017}, a power-law growth with disorder-dependent exponents is the signature of the RBCM domain growth kinetics for a quench in the LRO phase (within the simulation time-scales).

(c) A quench in the QLRO regime ($T_c^2<T<T_c^1$) is defined by interpenetrating domains with rough domain interfaces and disorder independent scaling function. Effect of $\epsilon$ on domain growth kinetics is weaker than the quench in the LRO regime, but the growth law is best described again by a power-law growth with disorder-dependent exponents on the time-scale of our simulations. 

(d) A brief exploration of the domain growth kinetics in bond-diluted clock model shows features which are qualitatively similar to the coarsening dynamics in RBIM.

This paper is organized as follows. In Sec. \ref{s1}, we discuss the model and present details of our numerical simulations. In Sec. \ref{result}, we present detailed numerical results from our simulations of the RBCM. Finally, in Sec. \ref{summary}, we conclude this paper with a summary and discussion of the results.


\section{Modeling and Simulation Details}
\label{s1}
\subsection{Random bond $q$-state clock model (RBCM)}
\label{ss1}

The model describes an ensemble of spins defined on a two-dimensional square lattice of size $N = L \times L$, where $L$ is the linear system size. Each site has four nearest neighbors and the lattice is connected via periodic boundary conditions along the $x$ and $y$ directions. In the $q$-state clock model, the spins are discrete and confined on the $xy$-plane where they can take $q$ discrete orientations specified by the equation
\begin{equation}
\label{refN}
\theta_n= \frac{2\pi n}{q},
\end{equation}
where $n$ = 0, 1, 2, ...., ($q$ - 1) denotes the discrete states of the spin. 
\begin{figure}[htbp]
\centering
\vskip 0.4 cm
\includegraphics[width=\columnwidth]{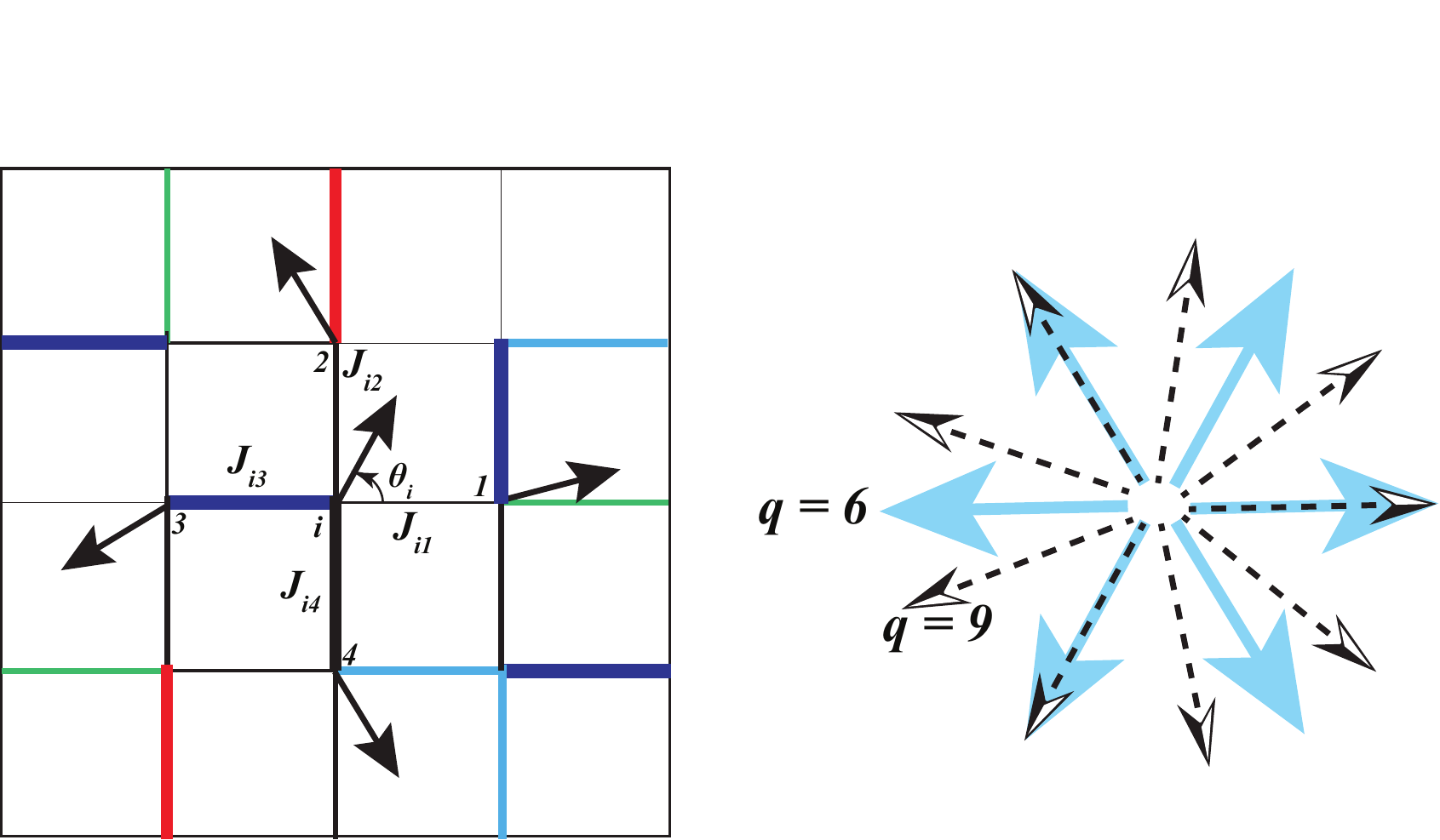}
\caption{(Color online) Schematic of RBCM showing random bond strength between the nearest neighbor sites (left) and possible orientations of spin vectors for $q=6$ and 9 (right). $1$, $2$, $3$, and $4$ represent the four nearest neighbors of site $i$. Different bond widths signify varying strength of bond-randomness.}
\label{fig1}
\end{figure}
The Hamiltonian for the $q$-state RBCM is defined as follows$\colon$
\begin{equation}
\label{refH}
H= -\sum\limits_{\langle ij\rangle} J_{ij} \vec{\sigma}_i\cdot \vec{\sigma}_j =-\sum\limits_{\langle ij \rangle} J_{ij} \cos (\theta_i - \theta_j) ,
\end{equation} 
where $\langle ij \rangle$ denotes summation over the nearest neighbors. $\vec{\sigma}_i=(\cos {\theta}_i, \sin {\theta}_i)$ denotes the unit vector representing the orientation of the spin at site $i$. Ferromagnetic coupling $\{J_{ij}\}>0$ between the two nearest neighbor sites $i$, $j$ is picked randomly from a uniform distribution $\in$ [1 - $\epsilon$/2, 1 + $\epsilon$/2], where the disorder amplitude $\epsilon$ describes a pure clock model ($\epsilon$ = 0) or a fully disordered clock model ($\epsilon$ = 2). A schematic diagram of this arrangement is shown in Fig.~\ref{fig1}.

\subsection{Simulation details for studying transition temperatures $T_c^1$ \& $T_c^2$}
\label{ss2}
Prior to investigating the phase ordering kinetics in RBCM, we quantify the deviation of the transition temperatures $T_c^1 (q,\epsilon=0)$ and $T_c^2(q,\epsilon=0)$ of pure clock model due to a finite effect of $\epsilon$. This characterization of the transition temperatures as a function of the bond randomness $\epsilon$ would help to locate the quench temperature in the LRO and QLRO regimes. 

To characterize $T_c^1$ and $T_c^2$, we make use of the canonical sampling Monte Carlo (MC) method with Wolff single cluster flipping algorithm \cite{wolff89} to equilibrate the system. A single Monte Carlo step (MCS) can be described as follows$\colon$

(a) A random mirror line with the normal vector $\vec{r} =(\cos {\delta},\sin {\delta})$ is chosen, where ${\delta}$ is a random discrete angle in the $xy$ plane \cite{Wu2012}. For even $q$, $\delta=n(\frac{\pi}{q})$, while for odd $q$, $\delta=(n+\frac{1}{2})(\frac{\pi}{q})$, where $n=0$, 1, 2, ...., $2q-1$. 

(b) A random site $i$ out of $N$ lattice sites is chosen for reflection of the spin $\vec{\sigma}_i =(\hat{x}\cos \theta_i + \hat{y}\sin \theta_i$) as follows \cite{wolff89}$\colon$ 
\begin{equation}
\label{refR}
\mathcal{R}(\vec{r})\vec{\sigma}_i=\vec{\sigma}_i-2(\vec{\sigma}_i\cdot\vec{r})\vec{r},
\end{equation}
Simplifying this equation we have the phase angle $\theta^\prime_i$ of the reflected spin $\vec{\sigma}_i$ as
\begin{equation}
\label{refR1}
\theta^\prime_i = \pi-\theta_i+2\delta_i
\end{equation}

(c) Nearest-neighbor site $j$ of site $i$ is added to the spin cluster according to the probability $\mathcal{P}$ \cite{wolff89}$\colon$
\begin{equation}
\label{refP}
\mathcal{P}(\vec{\sigma}_i, \vec{\sigma}_j)= 1-\exp (min [0, 2 \beta J_{ij}(\vec{r} \cdot \vec{\sigma}_i)(\vec{r} \cdot \vec{\sigma}_j)]),
\end{equation}
Simplifying Eq. \ref{refP} we get the probability as:
\begin{equation}
\mathcal{P}(\theta,\delta) = \cos (\theta_i-\delta)\cos (\theta_j-\delta),
\end{equation} 
where $\beta={1}/{k_BT}$ is the inverse temperature, $k_B$ is the Boltzmann constant and in simulations taken as unity.

(d) The cluster is then updated by reflecting all the spins about the line perpendicular to the normal vector $\vec{r}$. One Monte Carlo step (MCS) corresponds to $N$ such updates.

Upon reaching the equilibrium, various useful thermodynamic quantities such as magnetization ($m$), specific heat ($C_v$), ratio of the equilibrium correlation function ($g$) are computed. The magnetic order parameter $m$ is given by$\colon$
\begin{equation}
m=\frac{1}{N}\sqrt{s_x^2+s_y^2} ,
\end{equation}
where $s_x=\sum_{i=1}^{N}{\cos \theta_{i}}$, $s_y=\sum_{i=1}^{N}{\sin \theta_{i}}$, and $N=L^2$. 

Specific heat ($C_v$) per spin can be extracted from the fluctuations of the energy $E$ per spin$\colon$
\begin{equation}
C_v=\frac{1}{Nk_BT^2}[\langle E^2 \rangle-\langle E \rangle^2]
\end{equation}
The transition temperature $T_c^1$ is computed from Binder's fourth order cumulant of the order parameter $U_4$ \cite{binder2005} evaluated with respect to temperature. $U_4$ is defined as$\colon$
\begin{equation}
\label{BC}
U_4= 1-\frac{\langle m^4 \rangle}{3\langle m^2 \rangle ^2}. 
\end{equation}
The intersection points of the $U_4$ versus $T$ curves for different $L$ can precisely quantify $T_c^1$. This definition of $U_4$ can only characterize $T_c^1$ \cite{swarna2018}. In order to quantify $T_c^2$, similar to $U_4$, we define another cumulant $U_m$ \cite{baek2009} as: 
\begin{equation}
\label{um}
U_m=1-\frac{\langle m_\phi^4 \rangle}{2\langle m_\phi^2 \rangle ^2}, 
\end{equation}
where, $m_\phi=\langle \cos(q\phi) \rangle$, and $\phi=\tan^{-1}{\big(\frac{s_y}{s_x}\big)}$. Analogous to $U_4$, $U_m$ versus $T$ can accurately measure $T_c^2$. The ratio of the equilibrium magnetic correlation functions $g$ (see Section~\ref{ss3} for the definition of correlation function) which is defined as follows:
\begin{equation}
g=\frac{C(L/2)}{C(L/4)},
\end{equation}
(computed at two fixed distances $L/2$ and $L/4$, $L$ is the linear lattice size) can also provide good estimations of the transition temperatures for the dual phase transition \cite{surungan2005,tomita2002}. When plotted against $T$, the various isolated $g$ curves corresponding to different $L$ merge at a higher $T$ which signifies $T_c^1$ and then segregate again at a lower $T$ which marks $T_c^2$. 

\subsection{Simulation details for studying ordering kinetics}
\label{ss3}
Initial configuration of the system is prepared homogeneous assigning random angles to the spins defined in Eq.~\eqref{refN}, followed by a rapid quench in the LRO ($T<T_c^2$) or QLRO ($T_c^2<T<T_c^1$) regime 
(separately) at time $t=0$. The system then evolves via local spin updating Metropolis algorithm \cite{metropolis53}. In a single Monte Carlo step (MCS), the $L^2$ spins are randomly chosen from the lattice and updated as follows$\colon$

(a) The local energy of a spin $\vec{\sigma_i}=(\cos \theta_i,\sin \theta_i)$ is calculated using Eq.~\eqref{refH}. 

(b) A random rotation $\phi$ is given to the spin $\vec{\sigma_i}$ with $\phi=\frac{2 \pi n}{q}$, $n$ = 1 to ($q$ - 1). 

(c) The local energy is calculated again and the difference of these two energies is stored in $\Delta\mathcal{H}$. The rotated configuration is accepted with a probability $\mathcal{W}$$\colon$
\[
\mathcal{W} =
\begin{cases}
\exp({-\beta\Delta\mathcal{H}}) & \text{for} \Delta\mathcal{H} > 0, \\
1 & \text{for} \Delta\mathcal{H} \leqslant 0,
\end{cases}
\]
The energy change $\Delta\mathcal{H}$, resulting from the rotation of the spin $\theta_i$ $\rightarrow$ ${\theta_i}^\prime$, is defined as$\colon$
\begin{equation}
\label{delH}
\Delta\mathcal{H}=\sum\limits_kJ_{ik}\left\lbrace \cos(\theta_i-\theta_k)-\cos({\theta_i}^\prime-\theta_k)\right\rbrace,
\end{equation}
where $k$ refers to the nearest neighbors of site $i$.

Here we emphasize that the Monte Carlo (MC) method exploited in this study is commonly used to characterize the domain growth kinetics. In the context of Ising, Clock, and Potts models, two types of MC dynamics are considered: (a) system with non-conserved order parameter evolved via the single-spin-flip Glauber dynamics \cite{glauber} and (b) systems with conserved order parameter (mimics the particle-hole exchange in lattice gas or exchange of ions in binary alloys) evolved via probabilistic spin-exchange Kawasaki dynamics \cite{kawasaki}. The RBCM is a Glauber model, where the order parameter is non-conserved and the heat bath induces fluctuations in the system via single-spin-flips. The Glauber model describes non-conserved kinetics because the spin-flip processes make the total magnetization time-dependent whereas total magnetization remains constant over time in the Kawasaki dynamics, which involves spin-exchange mechanism.

The ordering kinetics of the RBCM can be examined by measuring the characteristic length scale $R(t)$ from the time dependence of the correlation function $C(\vec{r},t)$. If a single length scale $R(t)$ exists, domain morphology does not change with time $t$, apart from a scale factor. Therefore, the order-parameter correlation function $C(\vec{r},t)$ exhibits a dynamical-scaling \cite{Bray94,puri-wadhawan,binder-stauffer} defined as:
\begin{equation} 
\label{refC}
\begin{split}
C(\vec{r},t) & = \frac{1}{N}\sum_{i=1}^{N} [\langle \vec{\sigma}_i(t) \boldsymbol{.} \vec{\sigma}_{i+\vec{r}}(t) \rangle - \langle \vec{\sigma}_i(t) \rangle \boldsymbol{.} \langle \vec{\sigma}_{i+\vec{r}}(t) \rangle]_{av} \\
& = g(r/R(t)),
\end{split}
\end{equation} 
Here $[...]_{av}$ indicates averaging over different independent realizations of the disorder and $\langle ... \rangle$ denotes averaging over thermal fluctuations. To estimate the average domain size $R(t)$, one can measure the distance for which $C(\vec{r},t)$, averaged over several independent realizations, decays to an arbitrary value (say 0.2) for the first time. The time-dependent structure factor $S(\vec{k},t)$, which is the Fourier transform of the real-space correlation function $C(\vec{r},t)$, also used frequently to probe domain growth. In fact, neutron or light scattering experiments probe $S(\vec{k},t)$ \cite{puri-wadhawan} :
\begin{equation}
\label{SF}
S({\vec{k}},t)=\int d\vec{r} e^{i\vec{k}\cdot\vec{r}}C({\vec{r}},t),
\end{equation}
where $\vec{k}$ is the wave-vector of the scattered beam. The corresponding dynamical scaling form for $S(\vec{k},t)$ is:
\begin{equation}
\label{SFScaling}
S(\vec{k},t)=R^df(kR),
\end{equation}
where $d$ (here $2$) refers to the dimensionality and $f(p)$ is a scaling function of the form:
\begin{equation}
\label{fp}
f(p)=\int d\vec{x} e^{i\vec{p}\cdot\vec{x}}g(x).
\end{equation} 
The scaling functions $g(x)$ and $f(p)$ can uniquely describe the architecture of the ordering system. In simulations, one usually attempts to obtain the functional forms of $g(x)$ and $f(p)$ defined in Eq.~\eqref{refC} and Eq.~\eqref{fp} respectively. Bray and Puri (BP) \cite{BP} and (independently) Toyoki (T) \cite{T} used a defect-dynamics approach to propose that the presence of $n$-component topological defects yields a power-law or $generalized$ $Porod$ $tail$ of the following form for the scaled structure factor:
\begin{equation}
f(p) \sim p^{-(d+n)}, p \to \infty.
\end{equation}
For the XY model, $n=2$ and for the Ising model, $n=1$. $n$ is not a well-defined quantity for clock model and depends on the defects which drive the ordering. For vortex driven growth, $n=2$ whereas for interface driven growth, $n=1$. 


\section{Numerical Results}
\label{result}
In this section, we present numerical results of the RBCM for bond disorder strength $\epsilon$ = 0, 0.5, 1, 1.5, and 2. In Section~\ref{ss4}, $T_c^1(\epsilon)$ indicating the passage from disordered homogeneous phase to QLRO phase and $T_c^2(\epsilon)$ indicating the transition from QLRO to LRO phase, are quantified. Knowing $T_c^1(\epsilon)$ and $T_c^2(\epsilon)$ as functions of the disorder strength $\epsilon$, coarsening in specific to a temperature quench located in the QLRO and LRO is clearly identified. The corresponding results are presented in Section~\ref{ss5}.

\subsection{Estimation of $T_c^1(\epsilon)$ and $T_c^2(\epsilon)$}
\label{ss4}
The $q$-state clock model with $q$ = 6 and 9 are simulated on square lattice with linear sizes $L$ = 32, 64, 96, 128 and 256. Starting from a homogeneous initial configuration which mimics the high temperature disorder phase, the system is subsequently equilibrated using the Wolff cluster update algorithm \cite{wolff89} for disorder amplitudes $\epsilon$ = 0 (pure system), 1, 1.5 and 2 (maximum disorder). To achieve better statistics, the system is equilibrated for $10^6$ MCS and then various thermodynamic quantities, such as, $m$, $m^2$, $m^4$, $C_v$, $g$, $m_\phi$, $m_\phi^2$, $m_\phi^4$ are thermally averaged up to $5 \times10^5$ MCS. Data obtained are further averaged over 100 independent initial spin configurations. 

\begin{figure}[htbp]
\centering
\includegraphics[width=\columnwidth]{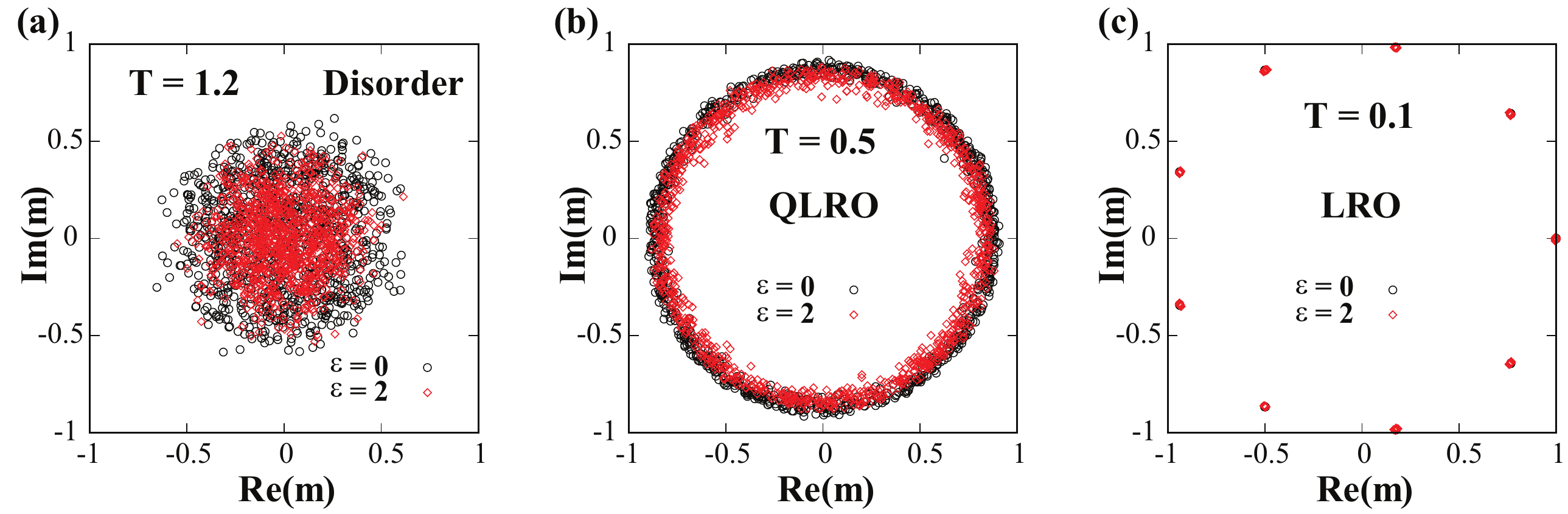}
\caption{(Color online) Distributions of the complex order parameter $m$ on the complex plane for $q$ = 9, obtained on a linear system size $L$ = 16, with 1000 ensembles, and $\epsilon$ = 0 (black open circle), 2 (red open rhombus). The system is cooled through various temperatures and corresponding distributions are recorded. (a) High temperature disordered phase at $T=1.2$. (b) Distribution at temperature $T=0.5$ signifies the QLRO phase and (c) LRO phase at $T=0.1$ are shown. The LRO phase displays nine degenerate ordering states for $q=9$. Effect of $\epsilon$ is visible in (a) and (b), where radii of the distributions decrease with increasing amplitude of the disorder.}
\label{fig2}
\end{figure}

In Fig.~\ref{fig2}, distribution of the order parameter $m=(s_x,s_y)$ on a complex plane are shown for 9-state clock model with $\epsilon$ = 0 (black open circle)  and 2 (red open rhombus), where, real part of $m$ is $Re(m)=s_x=\sum_{i=1}^{N} \cos \theta_i$ and imaginary part is $Im(m)=s_y=\sum_{i=1}^{N} \sin \theta_i$. Simulating over 1000 random initial configurations, data presented here for $L$ = 16 at three distinct temperatures ($T$), mark different phases: (a) $T$ = 1.2 (homogeneous disordered phase), (b) $T$ = 0.5 (QLRO phase) and (c) $T$ = 0.1 (LRO phase). These phases display (a) uniform distribution of spins at high $T$, where every spin in the lattice points to a random direction, (b) ring like distribution at intermediate $T$ signifying the Kosterlitz-Thouless (KT) \cite{kosterlitz74,thouless73} like phase, where spin waves and vortices arrange the spins and (c) nine isolated spots at low $T$ corresponding to the nine-fold degeneracy of the ferromagnetic ground state with equal probability for $q=9$. Since impurities tend to reduce the net magnetization, the radii of the distributions in (a) and (b) for $\epsilon=2$ are smaller compared to $\epsilon=0$.

\begin{figure}[htbp]
\centering
\includegraphics[width=\columnwidth]{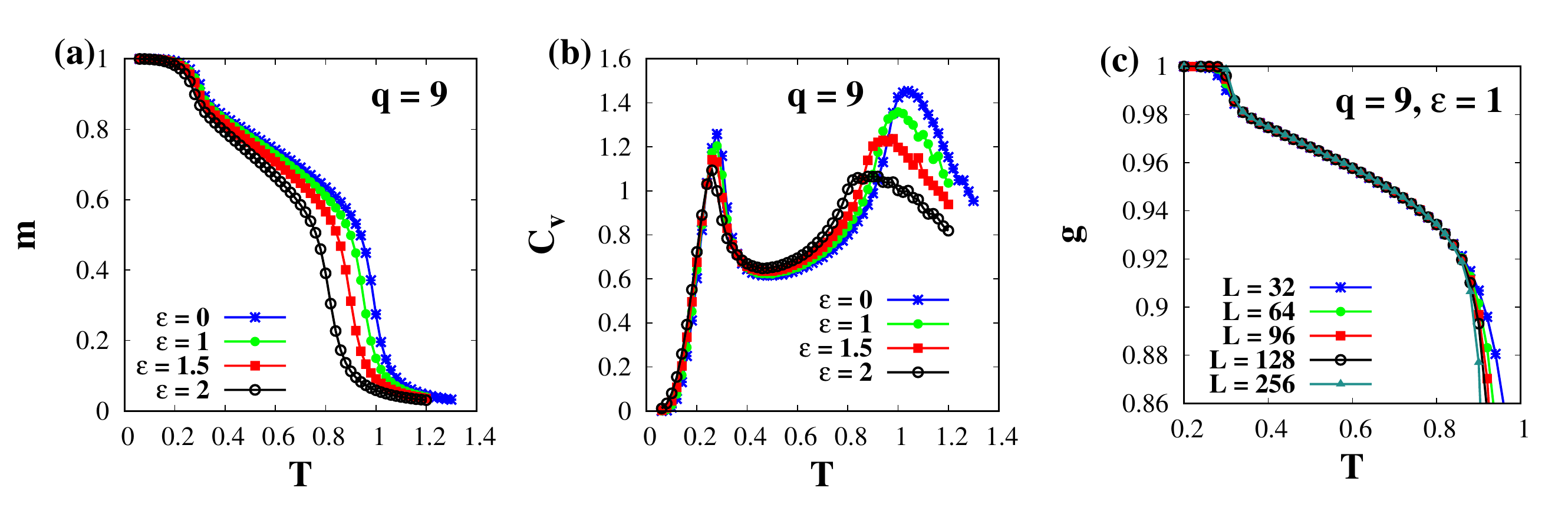}
\caption{(Color online) (a) Magnetization ($m$) versus temperature ($T$) for $q=$ 9, with $\epsilon$ = 0 (blue star), 1 (green solid circle), 1.5 (red solid square), and 2 (black open circle) and system size $L=128$. The two inflections correspond to two distinct phase transitions from disordered to QLRO at higher $T$ ($\sim T_c^1$) and QLRO to LRO at lower $T$ ($\sim T_c^2$). (b) Plot of specific heat $C_v$ versus $T$ shows dual peaks around $\sim T_c^1$ and $\sim T_c^2$ signifying phase transitions. Deviation of peaks as a function of $\epsilon$ is prominent near $T_c^1$ (disordered to QLRO) compared to $T_c^2$ (QLRO to LRO). (c) Ratio of equilibrium magnetic correlation functions $g = \frac{C(L/2)}{C(L/4)}$ for $\epsilon=1$ and $L=32$ (blue star), 64 (green solid circle), 96 (red solid square), 128 (black open circle), and 256 (teal solid triangle) also suggests similar inflections as shown in (a).}
\label{fig3}
\end{figure}

Demonstrating the effect of $\epsilon$ on the three phases of 9-state clock model, disorder dependency of the equilibrium thermodynamic parameters are quantified in Fig.~\ref{fig3}. The temperature dependency of (a) magnetization $m$, (b) specific heat $C_v$ and (c) ratio of equilibrium magnetic correlation functions $g$ against $\epsilon$ are depicted in Fig.~\ref{fig3}. $m$ versus $T$ for $\epsilon$ = 0 (blue star), 1 (green solid circle), 1.5 (red solid square), and 2 (black open circle) are shown in Fig.~\ref{fig3}(a) and is characterized by two distinct points (regions) of inflection: the inflection at high $T$ corresponds to a disordered to QLRO phase transition, while at low $T$ the inflection correlates with the QLRO to LRO phase transition. In fact, similar dual phase transition is reported in literature for $q \geqslant 5$ \cite{kadanoff77,swarna2018}. The points of inflection for $\epsilon>0$ are shifted toward smaller temperatures indicating phase transition temperatures decrease with the disorder strength. This scenario is further confirmed in the $C_v$ versus $T$ plot shown in Fig.~\ref{fig3}(b) where the peaks at higher $T$ (signifying disordered to QLRO phase transition) gradually shifted to lower temperature as the disorder amplitude $\epsilon$ is increased, whereas at low $T$ peaks (signifying QLRO to LRO phase transition), effect of $\epsilon$ appears to be marginal. This finding imply that $T_c^2$ is probably less affected by $\epsilon$ than $T_c^1$. We show the temperature dependence of magnetic correlation ratio $g=\frac{C(L/2)}{C(L/4)}$ \cite{tomita2002,surungan2005} for $L$ = 32 (blue star), 64 (green solid circle), 96 (red solid square), 128 (black open circle), 256 (teal solid triangle) and $\epsilon=1$. $C(L/2)$ and $C(L/4)$ are equilibrium correlation function of fixed length calculated using Eq.~\eqref{refC}. Characteristically, functional form of $g$ looks similar to $m$, with two major points of inflections. Although at high temperature data corresponding to different $L$ are separated, they gradually merge as the temperature is decreased toward the QLRO phase. Thus, merging of the data corresponds to $T_c^1$. The curves with different $L$ separate again at low enough temperature due to the discrete symmetry of the clock model and the point of separation corresponds to $T_c^2$. 

\begin{figure}[htbp]
\centering
\includegraphics[width=\columnwidth]{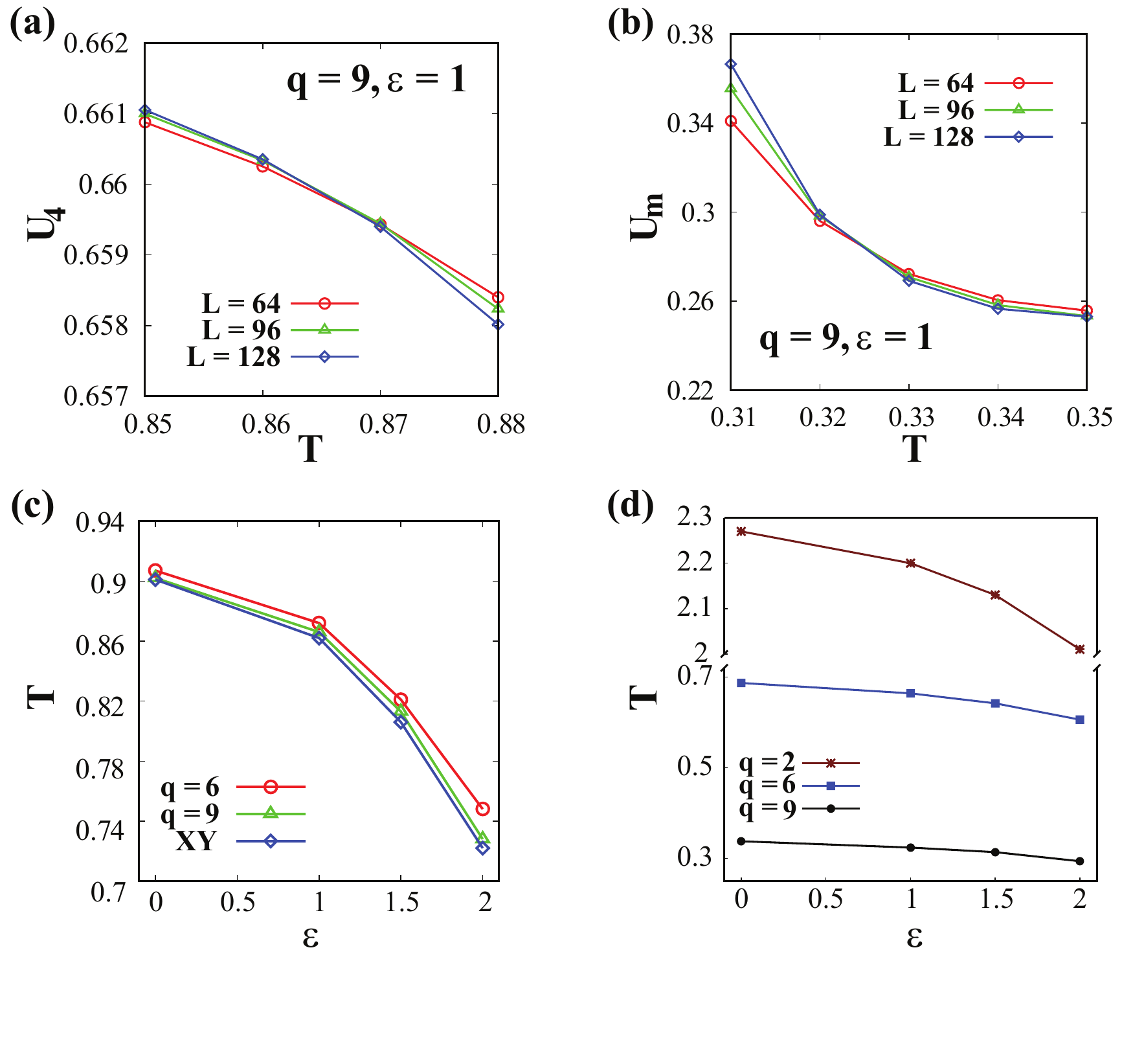}
\caption{(Color online) (a) $T_c^1(q=9, \epsilon=1)$ is quantified from the intersection of $U_4$ versus $T$ curves for $L$ = 64 (red open circle), 96 (green open triangle), and 128 (blue open rhombus). (b) $T_c^2(q=9, \epsilon=1)$ is extracted from the $U_m$ versus $T$ curves. (c) Change of transition temperatures with $\epsilon$ for various $q$, $q$ = 6 (red open circle), 9 (green open triangle), and XY (blue open rhombus) (disordered to QLRO phase). (d) $T_c^2$ (transition temperature, QLRO to LRO phase) as a decreasing function of $\epsilon$ for $q$ = 6- (blue solid square) and 9- (black solid circle) state clock model and compared with Ising model ($q=2$, maroon star, disordered to LRO phase).}
\label{fig4}
\end{figure}

Although it is possible to estimate the transition temperatures from Fig.~\ref{fig3}, more precise quantification of $T_c^1(\epsilon=1)$ and $T_c^2(\epsilon=1)$ can be made from the intersection of $U_4$ and $U_m$ curves for different $L$ as shown in Fig. \ref{fig4}(a) and Fig. \ref{fig4}(b). Effect of the bond randomness on transition temperatures is characterized in Fig.~\ref{fig4}(c). In this figure, data obtained for $q=6$ (red open circle), 9 (green open triangle), and $q=\infty$ (XY model) (blue open rhombus) depicts a decreasing of $T_c^1$ ($\sim 0.9$ to $\sim 0.7$) with $\epsilon$ in course of transition from disordered to QLRO phase; however, data for individual $q$ almost coincides with each other. Variation of the transition temperature with disorder from disordered to LRO phase ($T_c^2$) is shown in Fig.~\ref{fig4}(d), for $q=$ 6 (blue solid square), 9 (black solid circle) and compared with $q$ = 2 (Ising model, maroon star). Note that, data for XY model is omitted since it does not have a LRO regime. One recognizes that unlike $T_c^2$, $T_c^1$ in (a) varies significantly with $q$. Further, as $T_c^1$ and $T_c^2$ decrease with $\epsilon$, deviation in $T_c^1$ clearly dominates over the changes in $T_c^2$. This finding is analogous to the scenarios obtained for random bond Ising model and random bond XY model \cite{rp2005,manoj2017}. $T_c^1(\epsilon)$ and $T_c^2(\epsilon)$ corresponding to $q$ = 6 and 9 are tabulated in Table~\ref{table1}. As a plausible explanation for disorder affecting transition temperatures, one might consider interaction among the spins is perturbed due to random bond-strength between neighboring spins. In absence of disorder ($\epsilon=0$), $\{J_{ij}\} = 1$ and each spin interact with the adjacent spin uniformly across the system. When $\epsilon \neq 0$, $\{J_{ij}\}$'s are drawn from a uniform probability distribution [$1 - \frac{\epsilon}{2}, 1 + \frac{\epsilon}{2}$] with $\langle J_{ij} \rangle=1$, a spin no longer interacts with the adjacent spin uniformly. A weaker $\epsilon$ tend to reduce the probability of alignment between two neighboring spins, therefore, reducing the transition temperature. It turns out, that the overall effect of $\epsilon$ is not severe on $T_c^1$ and $T_c^2$ for bond-randomness as it is considered as a weak disorder unlike the disorder created by removing a site or bond from the lattice \cite{surungan2005}.

\begin{table}[htbp]
\centering
\caption{$T_c^1$ and $T_c^2$ as a function of $\epsilon$ for $q=6$ and $q=9$ state clock model.} 
\vspace{0.1in}
\label{table1}
\scalebox{1.1}{
\begin{tabular}{ | p{0.5 cm} | c | c | c | c | }
\hline
$\epsilon$ & $T_c^2(q=6)$ & $T_c^1(q=6)$ & $T_c^2(q=9)$ & $T_c^1(q=9)$ \\
\hline
0 & 0.687$\pm$0.002 & 0.907$\pm$0.001 & 0.338$\pm$0.002 & 0.902$\pm$0.002 \\
\hline
1 & 0.664$\pm$0.002 & 0.872$\pm$0.004 & 0.324$\pm$0.003 & 0.866$\pm$0.002 \\
\hline
1.5 & 0.642$\pm$0.001 & 0.821$\pm$0.002 & 0.314$\pm$0.002 & 0.813$\pm$0.003 \\
\hline
2 & 0.606$\pm$0.004 & 0.748$\pm$0.002 & 0.294$\pm$0.004 & 0.728$\pm$0.004 \\
\hline
\end{tabular}
}
\end{table}

\subsection{Phase ordering kinetics in random-bond $q$-state clock model}
\label{ss5}

In order to quantify the ordering kinetics in $q$-state clock model for $q$ = 6 and 9, we studied the evolution of spins on a square lattice of size $1024^2$ with periodic boundary conditions. Initially, all spins are randomly oriented as per Eq.~\ref{refN} mimicking the homogeneous phase at $T$ = $\infty$. Systems are independently quenched to (a) the LRO phase, $T<T_c^2(q,\epsilon)$ and (b) the QLRO phase, $T_c^2 (q,\epsilon)<T<T_c^1(q,\epsilon)$ [see Table~\ref{table1}] at $t=0$. Subsequently, spins are updated using the Metropolis algorithm up to $t=10^6$ MCS. All the statistical data presented here are averaged over 20 independent configurations of $\{J_{ij}\}$ and $\{\vec{\sigma_i}\}$. 

\begin{figure}[htbp]
\centering
\includegraphics[width=\columnwidth]{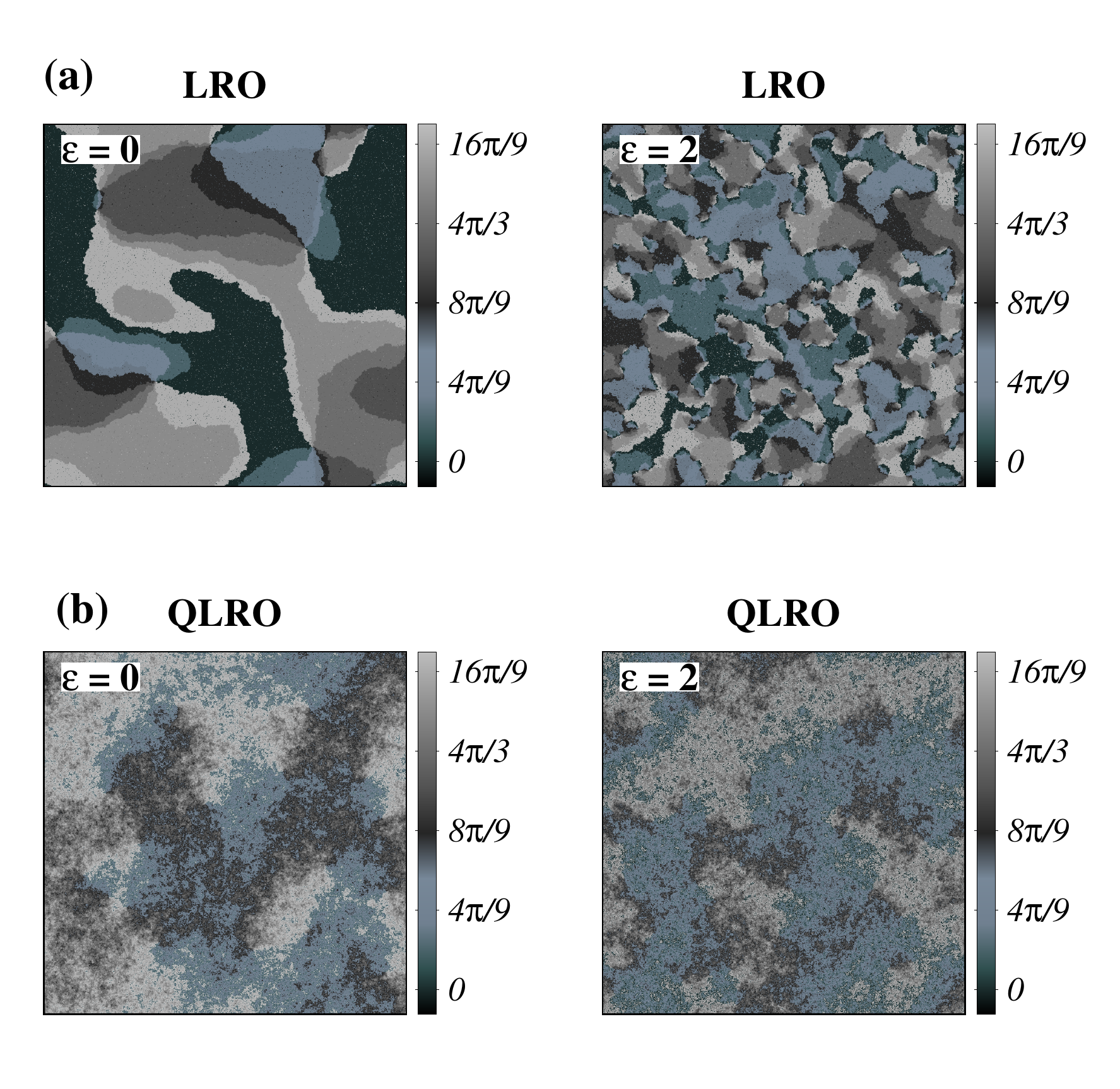}
\caption{Domain evolution snapshots for the 9-state clock model at $t = 10^5$ MCS after a quench from $T=\infty$ to the LRO regime (upper panel) and QLRO regime (lower panel) for $\epsilon=0$ and 2. The lattice size is $1024^2$. Shades specify different orientations of the $q=9$ clock spins according to Eq.~\eqref{refN}. (a) Quench in the LRO regime shrinks domain size considerably as $\epsilon$ increases. (b) In the QLRO regime, the effect of $\epsilon$ on the domain size is not apparent from the snapshots; however, domain boundaries are rough and interpenetrating.}
\label{fig5}
\end{figure}

Fig.~\ref{fig5} shows domain evolution snapshots of the 9-state clock model after a quench from $T$ = $\infty$ to the (a) LRO phase ($T<T_c^2$) and (b) QLRO phase ($T_c^2<T<T_c^1$) for $\epsilon$ = 0 and 2. Various shades represent domain orientations specified by Eq.~\ref{refN}. Distinct domains with sharp boundaries are the salient features of the LRO phase manifested in Fig.~\ref{fig5}(a), although a sharp decrease in the domain size is observed with a large $\epsilon$. Smaller domains in the latter scenario arise from the slow domain growth induced by bond randomness \cite{manoj2017}. Weaker bonds between neighboring spins impair their alignment. Although domains shrink at higher $\epsilon$, the domain morphologies are statistically similar and differ by a mere scale factor. In clock model, temperature quench in the LRO phase leads to two kinds of defects, domain walls, and point defects such as a vortex (net change in spin orientations surrounding the defect is +2$\pi$) or an antivortex (net change in spin orientations surrounding the defect is -2$\pi$). In the early stages of the domain evolution, the system coarsens via merging of domain walls (the well-defined domain boundaries we see in the snapshots of Fig.~\ref{fig5}(a)), as well as the annihilation of point defects with opposite topological charges $viz.$, vortices and antivortices; however, in the asymptotic limit, merging of domain walls becomes the dominant mechanism \cite{swarna2018}. Although energetically expensive interfaces and point defects are mostly eliminated from the system at the later stage of the coarsening of a pure system \cite{swarna2018}, snapshots at higher $\epsilon$ show many such defects with high energy barriers affecting the domain growth. A quench in the QLRO regime is characterized by interpenetrating and rough interfaces lacking compactness \cite{swarna2018}. Coarsening slows down at higher $\epsilon$, however, the effect of the disorder on domains cannot be ascertained from the morphologies shown in Fig.~\ref{fig5}(b). Characteristic length scale versus time for various $\epsilon$ in the QLRO phase can shed light on the domain growth kinetics.

\begin{figure}[htbp]
\centering
\includegraphics[width=\columnwidth]{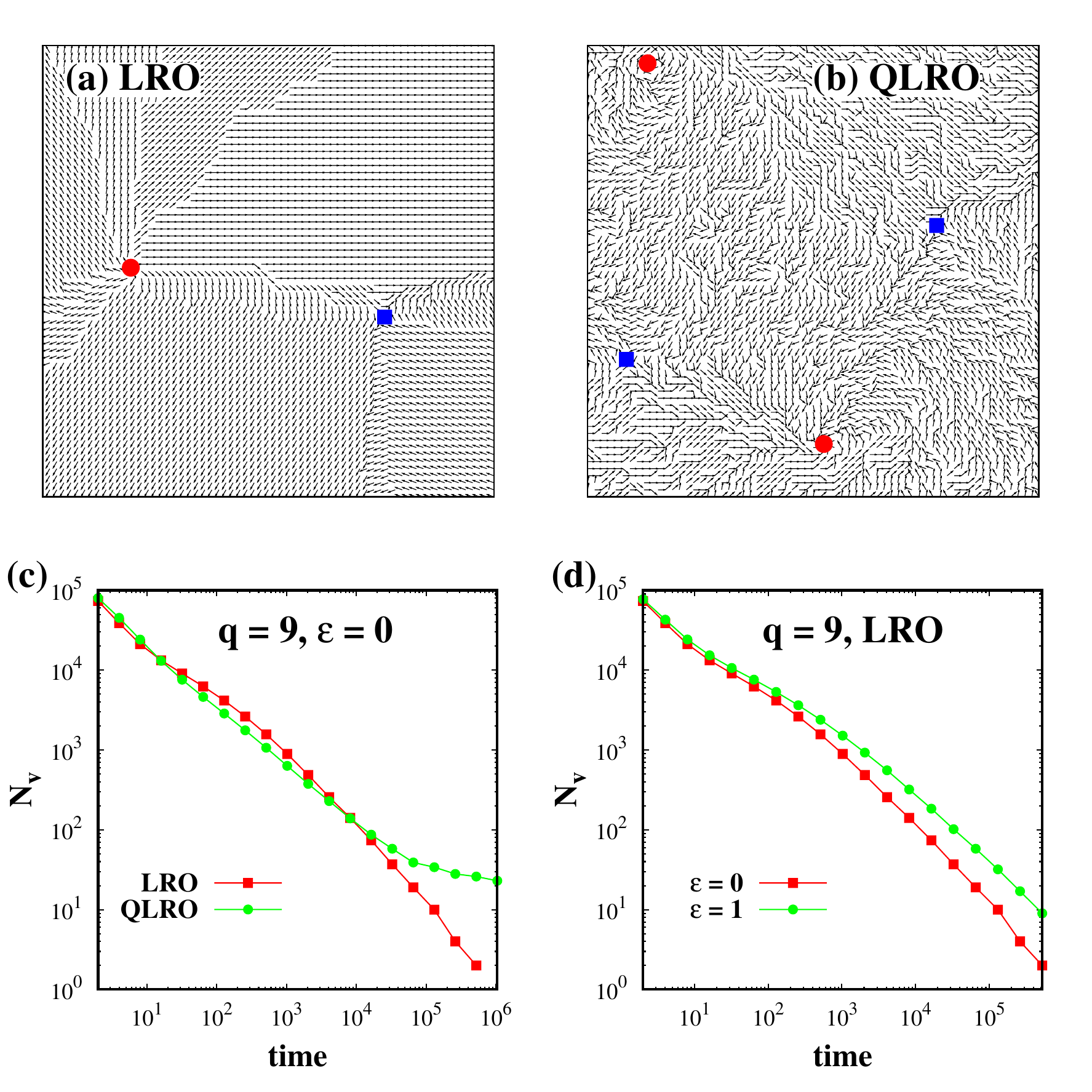}
\caption{(Color online) Vortex (red solid circle) and anti-vortex (blue solid square) from the simulation are shown for (a) LRO quench and (b) QLRO quench. For clarity, a $64^2$ corner is shown from a $1024^2$ lattice. The decay of vortices number ($N_v$) with time (on a log-log scale) are shown for (c) LRO (red solid square) and QLRO (green solid circle) regimes with $\epsilon=0$ and (d) in the LRO regime for different $\epsilon$, $\epsilon=0$ (red solid square) and $\epsilon=1$ (green solid circle). The anti-vortices number versus time plot is exactly similar to the number of vortices are always equal to the number of anti-vortices to keep the net topological charge neutral.}
\label{figg6}
\end{figure}

In Fig.~\ref{figg6}, we have shown images of vortex-antivortex pair in both (a) LRO and (b) QLRO regimes and vortex number $N_v$ as a decreasing function of time (on a log-log scale). Red solid circles represent vortices and blue solid squares represent anti-vortices. The snapshots are shown at $t=2^{15}$ and a $64^2$ corner of a $1024^2$ lattice is shown in each snapshot for better clarity. Fig.~\ref{figg6}(c) demonstrates a comparison of $N_v$ decay in the LRO and QLRO regime for a fixed disorder strength ($\epsilon=0$) whereas a comparative description of the decay of the point defects with $\epsilon$ is shown in Fig.~\ref{figg6}(d). In (c), we find $N_v$ decays faster with time in the LRO regime (red solid square) compared to the QLRO regime (green solid circle), although crossovers exist in the plot suggesting that the decay is non-monotonous. Besides, $N_v$ never reaches zero within the simulation time scale, nevertheless, for a quench in the LRO regime, it decays to a small number of defects which are expected to disappear at longer times. The effect of disorder on domain growth can be understood from the data in Fig.~\ref{figg6}(d) where $N_v(\epsilon=1)$ (green solid circle) is always higher than $N_v(\epsilon=0)$ (red solid square) signifying slower domain coarsening. 

Since a $q$-state clock model has $q$ equally favorable ground states, a domain interface can only emerge between two neighboring domains in $^qC_2=\frac{q(q-1)}{2}$ possible ways \cite{swarna2018}. But, in $d=2$ and $q\geqslant3$, three or more different domains can meet at a point and produce vortex or anti-vortex. For $q=9$, a quench in the LRO regime leads to nine different types of domains which would reach the final equilibrium state having a majority of the spins aligned along one of the nine directions. As the system runs toward the equilibrium, energetically expensive interfaces meet and coalesce to form larger domains and consequently point defects are also eliminated. Analyzing the time evolution snapshots (Fig.~\ref{figg6}(a) and Fig.~\ref{figg6}(b)), we find that vortices and antivortices are present as long as there are three or more different types of domains. The system will be completely devoid of the point defects in the very long time scale when only two different domains remain. While studying non-equilibrium dynamics, we normally do not reach that time scale and therefore we do not see a system where point defects are completely gone. 

A quench in the QLRO regime generates vortices and antivortices which is in sharp contrast to the domain wall interfaces observed in the LRO phase. Since the energy cost to create a vortex is higher than the corresponding energy cost of a domain interface, thermal fluctuation provided by the higher temperature in the QLRO regime can easily create such point defects. In other words, when the average angular fluctuation between the adjacent spins is large, domain walls are destabilized by the spin waves, resulting in vortices and antivortices in the QLRO regime \cite{swarna2018}. Thus, in the asymptotic limit, the merging of interfaces is the dominating mechanism of coarsening in the LRO regime, whereas in the QLRO regime, domain growth happens via the annihilation of the more energetic point defects.

It is worth mentioning that the LCA growth law $R(t) \sim t^{1/2}$ derived from the motion of the interfaces \cite{Bray94}, also describes the domain growth kinetics for clock model in the LRO regime \cite{swarna2018}. Nevertheless, the growth law we obtained for the clock model in the QLRO regime \cite{swarna2018} is valid for the $d=2$ XY model, $R(t) \sim (\frac{t}{\ln t})^{1/2}$. The $\ln t$ correction term in the denominator arises from the vortices present in the system. Our simulation data for the pure RBCM ($\epsilon=0$) are consistent with these theoretical predictions, suggesting that the diffusion of domain boundaries gives rise to the LCA growth law in the LRO regime and annihilating point defects prompts the XY-type growth law in the QLRO regime.

\begin{figure}[htbp]
\centering
\includegraphics[width=\columnwidth]{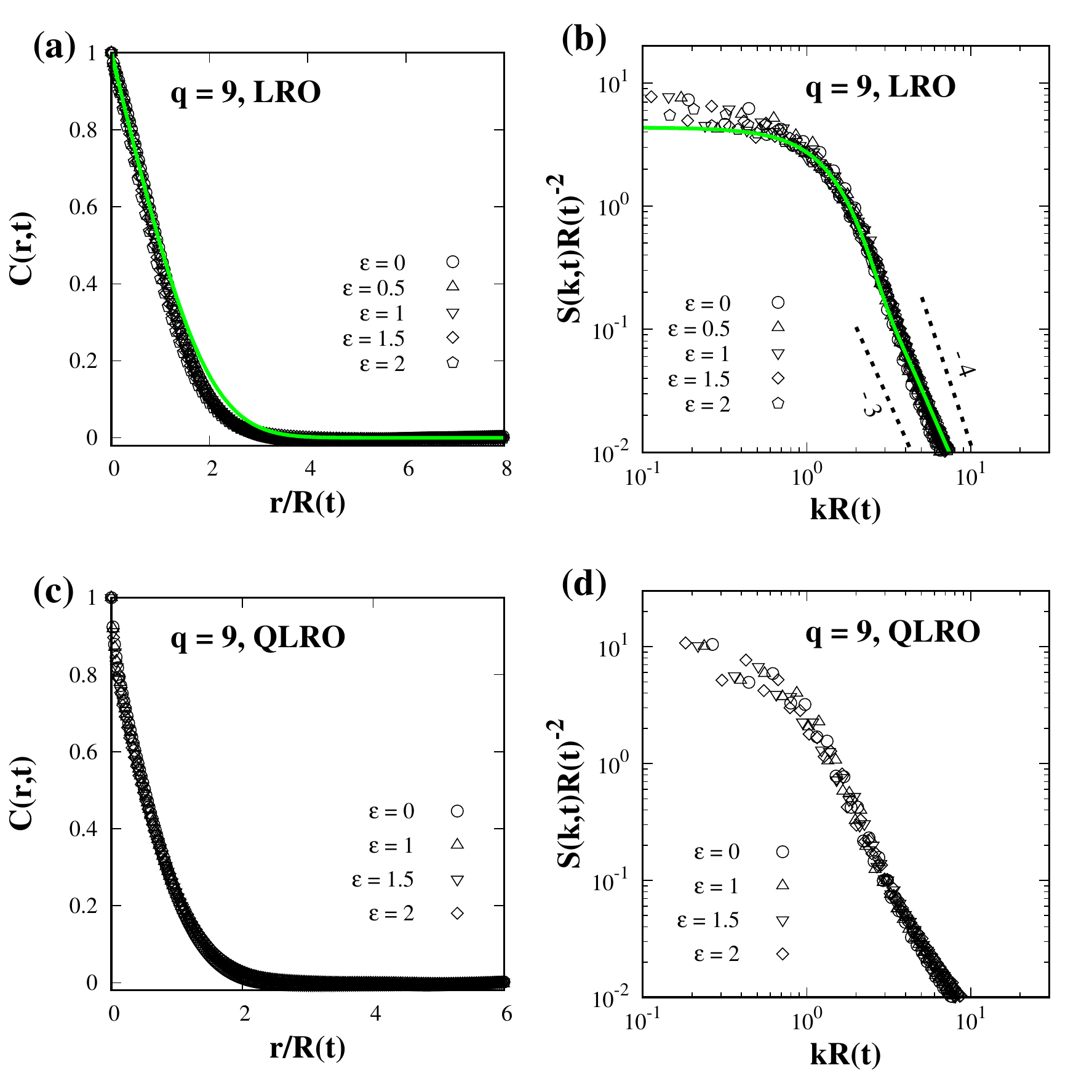}
\caption{Scaling of the correlation function and structure factor for 9-state clock model at $t = 10^4$ followed by a quench with $\epsilon$ = 0, 0.5, 1, 1.5, and 2. Data for the LRO regime shows (a) scaled $C(\textbf{r},t)$ versus $r/R(t)$ and (b) scaled structure factor (on a log-log scale) $S(k,t)R(t)^{-2}$ versus $kR(t)$ for a quench temperature $T=0.1$. The solid green (color online) curves in (a) and (b) signify the Bray-Puri-Toyoki (BPT) function for $n=2$ and it's Fourier transform respectively. The large-$k$ structure factor tail shows a slope $-3.263\pm0.021$, which lies between -3, slope for the Porod’s decay $S(k,t) \sim k^{-(d+1)}$, with $d=2$ and -4, slope for the generalized Porod’s law $S(k,t) \sim k^{-(d+n)}$ with $d=n=2$. Data for the QLRO regime shows (c) scaled correlation functions, $C(\textbf{r},t)$ versus $r/R(t)$, after a quench from $T=\infty$ to the QLRO regime ($T = 0.55$). (d) Scaled structure factor, $S(k,t)R(t)^{-2}$ versus $kR(t)$, for the data presented in (c).}
\label{fig6}
\end{figure}

In Fig.~\ref{fig6}, we demonstrate the dynamical scaling of the numerical data. The LRO regime [$T<T_c^2$] is quantified by plotting the scaled correlation function $C(r,t)$ versus $r/R(t)$ and the structure factor $S(k,t)R(t)^{-2}$ (Fourier transform of the correlation function) versus $kR(t)$ at a fixed $t=10^4$ MCS and for various $\epsilon$, as shown in Fig.~\ref{fig6}(a), (b), respectively. In addition, scaling of the correlation function and the structure factor holds good with respect to $t$ for a fixed value of the disorder amplitude $\epsilon$ (data not shown here). Our data convincingly establish that domain morphologies of the $q$-state clock model are statistically identical, independent of time and disorder amplitude (same scaling has also been established for $q=6$, data not shown here), as shown earlier in the Random-Bond Ising Model (RBIM) and Random-Bond XY Model (RBXYM) \cite{puri91,puri93,rp2005,manoj2017}. This feature is widely known as the Super-Universality (SU) in scaling. Physically, SU means that the domain morphologies are equivalent, regardless of the disorder amplitude. The SU property has been demonstrated extensively in literature for the spatial correlation function and structure factor in studies of non-conserved ordering kinetics \cite{puri91,puri93,bray-humayun,rp2004,rp2005,manoj2017,sicilia}. However, in the scaling of auto-correlation functions, recent studies \cite{puri2010,puri2011,puri2012,manoj2017} have shown clear dependence on the disorder amplitude and thus demonstrating that SU does not hold good for auto-correlation functions. For conserved dynamics, one observes a significant departure, where SU does not apply even for the spatial correlation function \cite{subir}. We further validate the scaling by fitting a green solid master curve (color online) on the data in Fig.~\ref{fig6}(a) known as the Bray-Puri-Toyoki (BPT) function for $n=2$ \cite{BP,T,puri-wadhawan} which have the following functional form:

\begin{equation}
\label{bpt}
f_{BPT}(r/R)=\frac{n\gamma}{2\pi}\left[B\left(\frac{n+1}{2},\frac{1}{2}\right)\right]^2 F\left(\frac{1}{2},\frac{1}{2};\frac{n+2}{2};\gamma^2\right),
\end{equation}
where $\gamma=\exp(-r^2/R^2)$, $B(x,y)\equiv \Gamma(x)\Gamma(y)/\Gamma(x+y)$ is the Euler's beta function,$F(a,b;c;z)$ is the hypergeometric function $_2F_1$, and $R$ is the average defect length-scale. The BPT result is valid for $n \leqslant d$, and corresponds to the cases where topological defects are present (XY model, clock model) \cite{puri-wadhawan}. The green solid curve (color online) fitted with the large-$k$ behavior of the structure factor tail in Fig.~\ref{fig6}(b) is the Fourier transform of the BPT function and shows a slope $-3.263 \pm 0.021$ (in a log-log scale) consistent with our earlier finding \cite{swarna2018}. The physical significance of the structure factor tail having a slope between $-3$ (Porod’s Decay, $S(k,t) \sim k^{-(d+1)}$) and $-4$ (Generalized Porod’s Law, $S(k,t) \sim k^{-(d+n)}$) \cite{Bray94,puri-wadhawan} lies in the fact that in $q$-state clock model domain growth involves both sharp domain interfaces and point vortices-antivortices as topological defects. Dynamical scaling after a quench in the QLRO phase [$T_c^2(q=9,\epsilon)<T<T_c^1(q=9,\epsilon)$] is shown in Fig.~\ref{fig6}(c) and Fig.~\ref{fig6}(d) for various $\epsilon$ and at $t=10^4$ MCS. Fig.~\ref{fig6}(c) shows the scaled data of $C(r,t)$ versus $r/R(t)$, whereas, Fig.~\ref{fig6}(d), shows the scaled data of $S(k,t)R(t)^{-2}$ versus $kR(t)$ (on a log-log scale). We further confirm that the scaling is valid for a fixed $\epsilon$ and different $t$ for $q=6$ (data not shown here). The data presented in (c) and (d) also establishes that domain architecture is independent of disorder for a quench in the QLRO regime and satisfy SU. The extracted slope from the large-$k$ behavior of the structure factor tail shown in Fig.~\ref{fig6}(d) is $-1.91\pm0.04$. As explained in our earlier communication \cite{swarna2018}, this non-integer slope is the suggestive of interpenetrating domains with rough domain interfaces (see Fig.~\ref{fig5}) and can be described as a non-Porod behavior. This type of non-Porod behavior is indicative of the scattering from the rough domain morphologies and has been observed in other statistical systems \cite{manoj2014,puri2016}.

Characterizing length scale $R(t)$ with time is crucial to understand the domain growth kinetics. In a pure ($\epsilon=0$) $q$-state clock model, the equation of motion for sharp domain interfaces (see Fig.~\ref{fig5} LRO), we consider $v(\vec{a})$ as the normal interfacial velocity along the $\hat{n}$-direction, where $\hat{n}$ is the unit vector normal to the interface and $\vec{a}$ is the tangent along the interface. The domain coarsening at the asymptotic limit is governed by the interface curvature, as the system approaches the equilibrium via energy dissipation and shrinking the surface area. For a curvature driven growth, the relation between interface motion and local curvature according to the Allen-Cahn \cite{puri-wadhawan} equation is: 
\begin{equation}
\label{AC}
v(\vec{a})=-\vec{\nabla}.\hat{n}=-K(\vec{a}),
\end{equation}
where $v \sim dR/dt$ and $K \sim 1/R$ denotes the local curvature of the interface. Upon integrating, Eq.~\ref{AC} yields the diffusive growth law, $R(t) \sim t^{1/2}$, known as the LCA (Lifshitz-Cahn-Allen) growth law and is valid for non-conserved systems. Domain growth in $q$-state clock model for a quench in the LRO phase ($T<T_c^2$) follows the LCA growth law \cite{Bray94,puri97,corberi2006}.

\begin{figure}[htbp]
\centering
\includegraphics[width=\columnwidth]{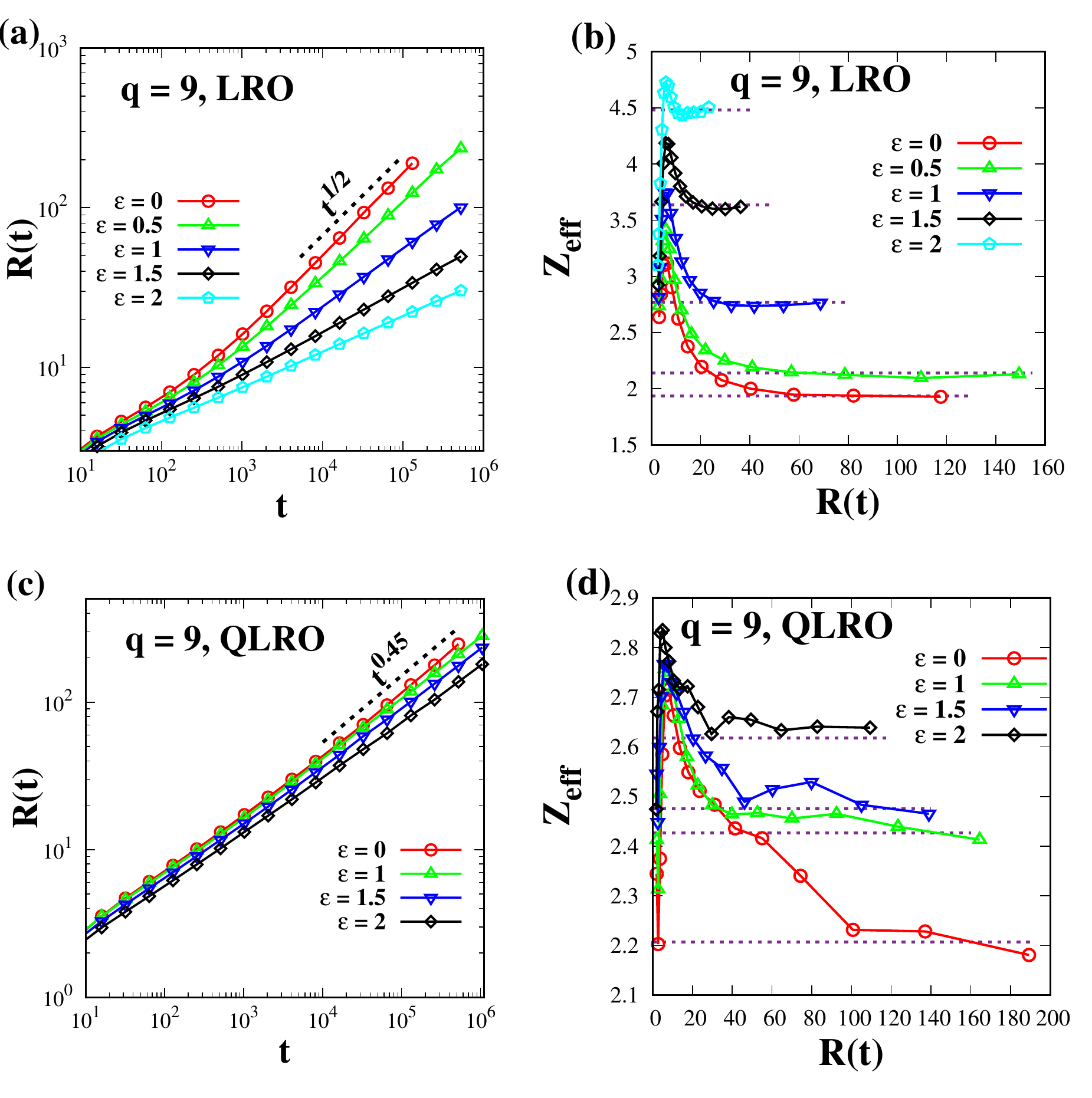}
\caption{(Color online) Length scale data for a quench in the LRO regime, $q=9$ and quench temperature $T=0.1$. (a) $R(t)$ versus $t$ (on a log-log scale) for $\epsilon$ = 0 (red open circle), 0.5 (green open triangle), 1 (blue open inverted triangle), 1.5 (black open rhombus) and 2 (cyan open pentagon). Dashed line indicates the growth law of the pure clock model $R(t) \sim t^{1/2}$. (b) $Z_{eff}$ versus $R(t)$ plot for the data sets in (a). The dashed lines imply $\bar{z}(\epsilon)=\psi_{LRO}(\epsilon)^{-1}$. (c) Plot of $R(t)$ versus $t$ for RBCM with $q=9$ and a quench in the QLRO regime for $\epsilon$ = 0 (red open circle), 1 (green open triangle), 1.5 (blue open inverted triangle), and 2 (black open rhombus). The growth law for the pure case ($\epsilon=0$) $R(t) \sim t^{0.45}$ (or equivalently $R(t) \sim (t/\ln t)^{1/2}$) is marked by the dashed line. (d) Effective exponent $Z_{eff}$ versus $R(t)$ for the plot in (c). The dashed lines represent $\bar{z}^\prime=\psi_{QLRO}(\epsilon)^{-1}$.}
\label{fig7}
\end{figure}

Length-scale data estimated from the domain configuration of $q=9$ state clock model are shown in Fig.~\ref{fig7}. Plotting $R(t)$ versus $t$ in the LRO regime on a log-log scale for different $\epsilon$ (see Fig.~\ref{fig7}(a)), growth kinetics can be illustrated by an algebraic law of the following form:
\begin{equation}
\label{GL}
R(t) \sim t^{\psi_{LRO}} \sim t^{1/\bar{z}}
\end{equation}
where $\psi_{LRO}(\epsilon) = 1/\bar{z}$ is a disorder-dependent exponent. Fitting a function $f(x)=ax^b$ with the simulation data for $\epsilon=0$ we extract the asymptotic growth exponent $\psi_{LRO}(\epsilon=0) \sim 0.5$, as indicated by the dashed line placed as a guide to the eye. We make two important observations from the data plotted in Fig.~\ref{fig7}(a): (a) pre-asymptotic growth crosses over to an asymptotic growth with higher exponent. In the clock model, this arises due to the pre-asymptotic domain growth governed by the merging of the domain walls and annihilating point defects, subsequently switching to a faster asymptotic domain growth solely driven by the merging of domain interfaces \cite{swarna2018}. (b) The data confirms an asymptotic algebraic growth of $R(t)$ as defined in Eq.~\ref{GL}. The disorder dependent growth exponents $\psi_{LRO}(\epsilon)$ extracted from the measurement of the corresponding slopes are tabulated in Table~\ref{table2}. 

In order to further investigate the nature of the domain growth presented in Fig.~\ref{fig7}(a), we have calculated the effective exponent $Z_{eff}(R)$ defined by:
\begin{equation}
\label{zeff}
\frac{1}{Z_{eff}} = \frac{d(\ln R)}{d(\ln t)}
\end{equation}
and plotted $Z_{eff}(R)$ versus $R(t)$ in Fig.~\ref{fig7}(b). Plateau observed for $\epsilon=0-1.5$ (red open circle, green open triangle, blue open inverted triangle and black open rhombus respectively), signifies the system manifesting a power-law growth described by Eq.~\ref{GL}. The power law growth is consistent with the earlier findings of RBIM and RBXYM \cite{rp2005,manoj2017,puri2010,puri2011,henkel2008}. For maximum disorder amplitude $\epsilon=2$ (cyan open pentagon), we notice a slight upward curvature at very later stage of the growth pointing to a slow coarsening. A probable reason could be the lack of activation energy required to conquer the energy barriers imposed by the disorder as the quench temperature $T=0.1$ $[T_c^2(\epsilon=2) \sim 0.294]$ does not providing enough fluctuation. A similar signature in the effective exponent plot was observed in RBIM and $d=3$ RBXYM \cite{puri2011,manoj2017}. The straight horizontal broken lines corresponding to every $\epsilon$-value represent $\bar{z}(\epsilon) = 1/\psi_{LRO}(\epsilon)$, where $\psi_{LRO}(\epsilon)$ are taken from Table~\ref{table2}. An inspection of the length scale data of $q= 4$ and $6$ also manifest disorder affected domain coarsening, see the Appendix for details. In contrast to our current findings, a crossover from algebraic to logarithmic growth was suggested for both RBIM \cite{puri2010,puri2011} and $d=3$ RBXYM \cite{manoj2017}, however, within our simulation regime the signature remains elusive. Perhaps, an extensive and large scale simulations are required to confirm a logarithmic growth. 

Data presented in Fig.~\ref{fig7}(a) and Fig.~\ref{fig7}(b), is insightful to explain domain growth affected by bond randomness. Real systems we observe in nature, are far from being pure and isotropic - there are typically two types of impurities real systems contain, annealed (mobile) and quenched (immobile) impurities. Generally, at early times when the length scales are small, the growing domains are not affected by the quenched disorder and growth laws for pure and disordered systems are the same. In the presence of disorder, bonds affected by the quenched disorder act as traps for domain boundaries and the energy barrier ($E_B$) depending on the domain size. Thus, at early times when the domain size and barriers are small, the coarsening dynamics are not affected by the disorder. At later times, domains become bigger and are trapped by disordered bonds hindering domain growth. Once a domain wall is trapped in a metastable state, only thermal activation can move it over the corresponding energy barrier. Thus, thermal fluctuations drive asymptotic domain growth in disordered systems, unlike the pure system where thermal fluctuations are irrelevant (data not presented here). We infer that the presence of energy barriers arising from induced disorders, slow down the coarsening process. The form of $E_B$ could be discussed from our previous understanding of domain growth in RBIM \cite{rp2005,puri2010,puri2011}. It has been initially argued that $E_B$ scales logarithmically with $R(t)$ having the following barrier-scaling form, $E_B \sim \epsilon\ln(1+R)$ and $R(t)$ obeys an algebraic growth law with disorder-dependent exponent \cite{rp2005}. Further investigations \cite{puri2010,puri2011} suggest that $E_B$ scales as $E_B \sim \epsilon R^\gamma$ \cite{HH}, where $\gamma$ is the barrier exponent, and yields a logarithmic growth in the asymptotic limit, $R(t) \sim (\ln t)^{1/\gamma}$, succeeding the regime of the algebraic domain growth.

In Fig.~\ref{fig7}(c), (d), we discuss the coarsening phenomena in $q$ = 9 after the system is rapidly quenched from a homogeneous initial configuration at $T=\infty$ to 0.55 ($T_c^2<T<T_c^1$) in the QLRO regime for $\epsilon=0$ (red open circle), 1 (green open triangle), 1.5 (blue open inverted triangle) and 2 (black open rhombus). In a previous communication \cite{swarna2018}, and in Fig.~\ref{fig5}(b), we find that the domain morphologies in the QLRO regime can be best described by interpenetrating domains lacking compactness and the domains are also devoid of well-defined domain interfaces. For such domain architecture, we argued that, phase ordering kinetics proceeds via annihilation of point defects (vortices and antivortices) \cite{swarna2018}. In pure 9-state clock model ($\epsilon=0$), domain growth law in the QLRO regime can be described by the pure XY model growth law in $d=2$, $R(t) \sim (t/\ln t)^{1/2}$ \cite{swarna2018}. For RBCM with $q=9$ (QLRO quench), we expect that the governing growth law would follow the similar kind of growth we just experienced in the LRO scenario but with different growth exponents due to contrasting domain morphologies:
\begin{equation}
\label{GL2}
R(t) \sim t^{\psi_{QLRO}} \sim t^{1/\bar{z}^{\prime}},
\end{equation}
where $\psi_{QLRO} = 1/\bar{z}^{\prime}$ is the disorder-dependent growth exponent for a quench in the QLRO regime. Fig.~\ref{fig7}(c) shows the plot of $R(t)$ versus $t$ and in the asymptotic limit, domain growth for the pure case ($\epsilon=0$) can be described by $R(t) \sim t^{0.45}$ as indicated by the dashed line. This plot with a logarithmic correction of the time $t$ retrieves the pure XY model growth law mentioned above which is also the governing growth law of PCM with $q=9$ and QLRO quench \cite{swarna2018}. $\psi_{QLRO}$ for various $\epsilon$ tabulated in Table~\ref{table2}, signify moderate effect of the disorder on the growth process as visible from the domain morphologies displayed in Fig.~\ref{fig5} (b). Fig.~\ref{fig7}(d) shows the effective exponent $Z_{eff}$ versus $R(t)$ corresponding to the data set in Fig.~\ref{fig7}(c) where $Z_{eff}$ is defined in Eq.~\eqref{zeff}. Although an extended plateau, observed in the LRO regime (Fig.~\ref{fig7}(b)), is absent in the QLRO regime, the data supports a power-law domain growth with disorder-dependent exponents which validates Eq.~\ref{GL2}. We stress upon the fact that a larger quench temperature in the QLRO phase plays a significant role in the outcome of Fig.~\ref{fig7}(d) where we could not reach the smooth plateaus as we observe for the LRO in Fig.~\ref{fig7}(b). To increase the quality of data one needs to average over a significantly large number of independent realizations which is computationally very expensive due to limited resources. The dashed horizontal lines in represent $\bar{z}^\prime = 1/\psi_{QLRO}(\epsilon)$, where $\psi_{QLRO}(\epsilon)$ are taken from Table~\ref{table2}. For our analysis of the domain growth in $q=6$ state RBCM after a QLRO quench, see the Appendix.

It is worth mentioning that the power-law growth observed in the ordering kinetics of RBCM fits well within the time-scales of our simulation. As seen in RBIM \cite{puri2011,puri2012} and $d=3$ RBXYM \cite{manoj2017}, we may expect a crossover of the growth law from the present algebraic regime to a slower logarithmic regime, asymptotically. 

\begin{figure}[htbp]
\centering
\includegraphics[width=\columnwidth]{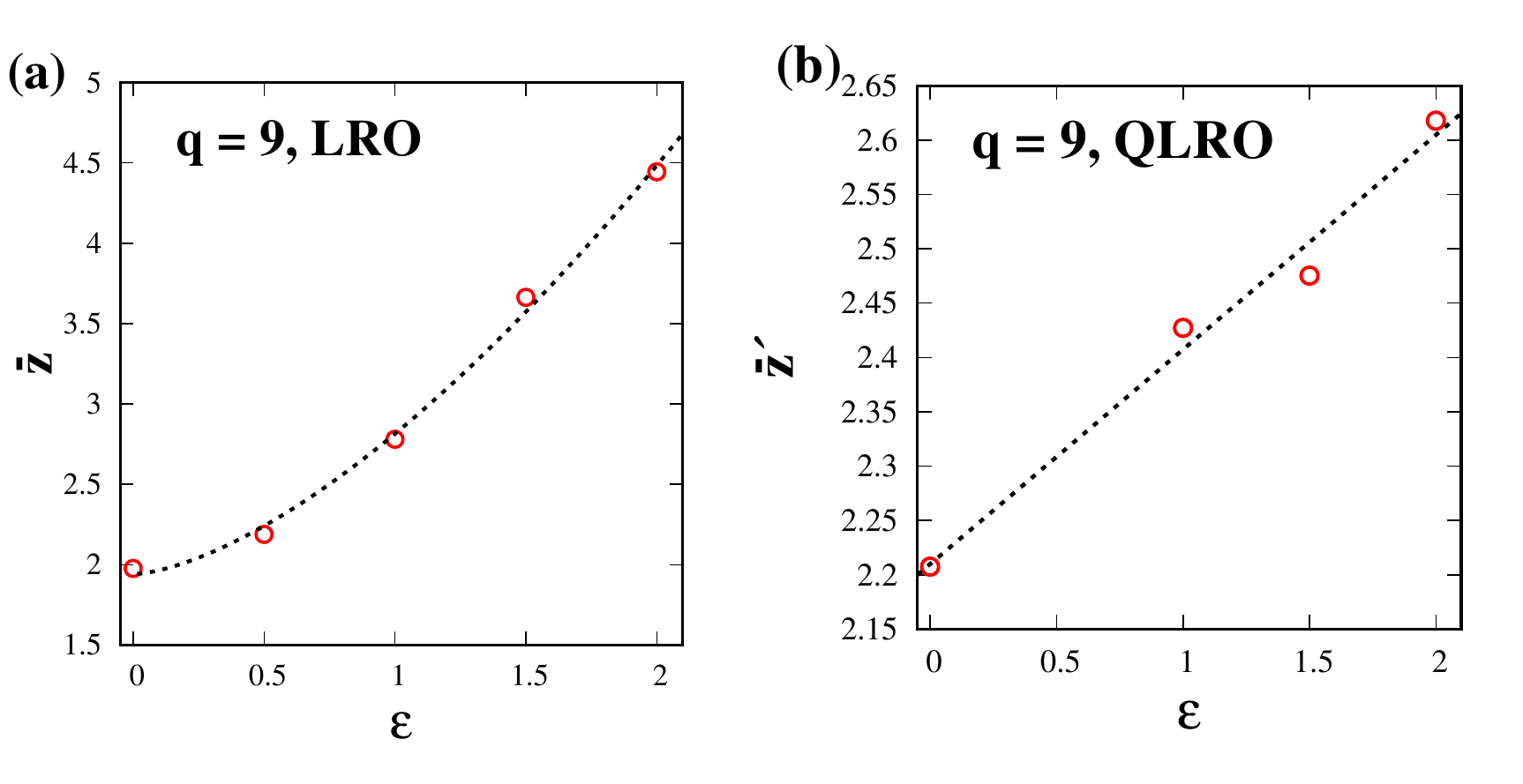}
\caption{(Color online) (a) $\bar{z}(\epsilon)$ (extracted from Fig~\ref{fig7}(b)) versus $\epsilon$ for quench in the LRO regime. The dashed line is a power-law fit: $\bar{z}(\epsilon)=1.94+0.874\epsilon^{1.54}$. (b) $\bar{z}^\prime(\epsilon)$ (extracted from Fig~\ref{fig7}(d)) is plotted against $\epsilon$ for quench in the QLRO regime. The dashed straight line represents the best fit: $\bar{z}^\prime(\epsilon)=0.197\epsilon+2.209$.}
\label{fig8}
\end{figure}

Exponents $\bar{z}=1/\psi_{LRO}(\epsilon)$ extracted from Fig.~\ref{fig7}(b) are plotted in Fig.~\ref{fig8}(a) with $\epsilon$. Paul $et~al.$ earlier argued that $\bar{z}$ scales linearly with $\epsilon$ \cite{rp2004,rp2005}. We observe that $\bar{z}$ increases non-linearly with $\epsilon$ as reported in the earlier studies of RBIM \cite{henkel2006,henkel2008} and RBXYM \cite{manoj2017} and can be fitted with function:
\begin{equation}
\bar{z}=\lambda+k\epsilon^{\alpha}
\end{equation}
where $\lambda$, $k$, and $\alpha$ are the fitting parameters. The best fit is achieved by setting $\lambda = 1.94$, $k=0.874$, $\alpha=1.54$. For a quench in the QLRO phase, $\bar{z}^\prime(\epsilon)$ are extracted from Fig.~\ref{fig7}(d). To analyze how $\bar{z}^\prime$ behaves with $\epsilon$, we plot $\bar{z}^\prime$ versus $\epsilon$ in Fig.~\ref{fig8}(b) and found a linear fit. A best fit to the data suggests, $\bar{z}^\prime=0.197\epsilon+2.209$. A qualitative comparison between Fig.~\ref{fig8}(a) and Fig.~\ref{fig8}(b) implies that domain growth of RBCM for a quench in the LRO regime has been affected more severely in presence of disorder, compared to the quench in the QLRO phase.

\begin{table}[htbp]
\centering
\caption{Growth exponents $\psi_{LRO}(\epsilon)$ and $\psi_{QLRO}(\epsilon)$ for $q$ = 9.}
\vspace{0.1in} 
\label{table2}
\scalebox{1.1}{
\begin{tabular}{ | p{1.0 cm} | c | c |}
\hline
$\epsilon$ & $\psi_{LRO}(\epsilon)$ & $\psi_{QLRO}(\epsilon)$ \\
\hline
0 & 0.506 $\pm$ 0.003 & 0.453 $\pm$ 0.004 \\
\hline
1 & 0.359 $\pm$ 0.003 & 0.412 $\pm$ 0.002 \\
\hline
1.5 & 0.273 $\pm$ 0.001 & 0.404 $\pm$ 0.002 \\
\hline
2 & 0.225 $\pm$ 0.001 & 0.383 $\pm$ 0.002\\
\hline
\end{tabular}
}
\end{table}

\subsection{Ordering kinetics in bond-diluted $q$-state clock model (BDCM)} \label{ss6}
Bond randomness in a lattice model can also be implemented by depleting the interaction between neighboring spins. Here we briefly discuss our findings of the $q$-state bond-diluted clock model (BDCM) where the Hamiltonian is represented by Eq.~\eqref{refH}, but $J_{ij}$ obeys the following distribution:
\begin{equation}
P(J_{ij})=p_{bond}\delta(J_{ij}-J)+(1-p_{bond})\delta(J_{ij}).
\end{equation}
$p_{bond}$ is the concentration of existing bonds and $J=1$ is the ferromagnetic coupling constant. For this investigation, we have taken $q=6$ and simulate the system on a two-dimensional square lattice of linear dimension $L=512$. Followed by a rapid quench from the high temperature ($T \to \infty$) homogeneous phase to a temperature $T$, where (a) $T<T_c^2(p_{bond})$ and (b) $T_c^2(p_{bond})<T<T_c^1(p_{bond})$ we simulate up to a maximum time $t=10^6$. Transition temperatures $T_c^1(p_{bond})$ and $T_c^2(p_{bond})$ for $q=6$ are taken from an earlier study by Surungan $et$ $al.$ \cite{surungan2005}. The simulations are done for various bond concentrations, from the pure case $p_{bond}=1$ to $p_{bond}=0.7$ and for each concentration data are averaged over $20$ independent realizations of spin configurations and bond distribution $P(J_{ij})$.

\begin{figure}[htbp]
\centering
\includegraphics[width=\columnwidth]{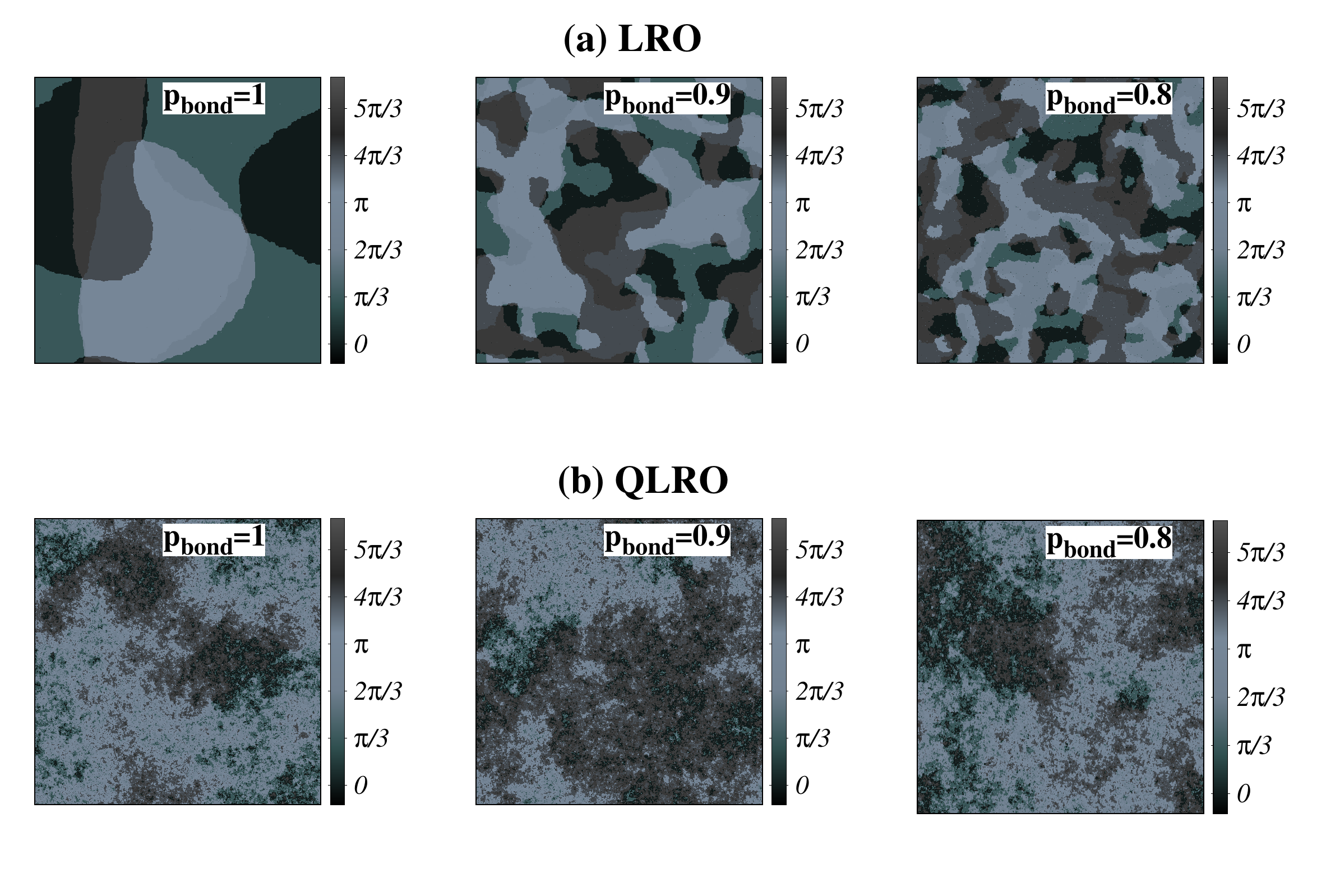}
\caption{Domain morphologies of the 6-state BDCM at $t = 10^5$ MCS after a quench from $T=\infty$ to the (a) LRO regime and (b) QLRO regime for $p_{bond}=1$, 0.9, and 0.8. The size of the simulation box is $512^2$. Gray color shades in the colorbar represent the six different orientations of the $q=6$ clock spins.}
\label{fig9}
\end{figure}

Domain morphologies in the BDCM after a quench in the LRO and QLRO regimes are demonstrated in Fig.~\ref{fig9}(a) and Fig.~\ref{fig9}(b) respectively at $t=10^5$ for $p_{bond} = 1, 0.9$, and 0.8 where for a quench in the LRO regime, we see that domain sizes have been greatly affected by the quenched bond-dilution but again in the QLRO regime, this effect is indistinguishable. The domain evolution snapshots in BDCM are qualitatively very similar to the RBCM snapshots shown in Fig.~\ref{fig5} where sharp domain interfaces with well-defined domain boundaries are signatures of the LRO phase, whereas rough interpenetrating domains with no precise domain boundaries are the salient features of the QLRO phase. 

\begin{figure}[htbp]
\centering
\includegraphics[width=\columnwidth]{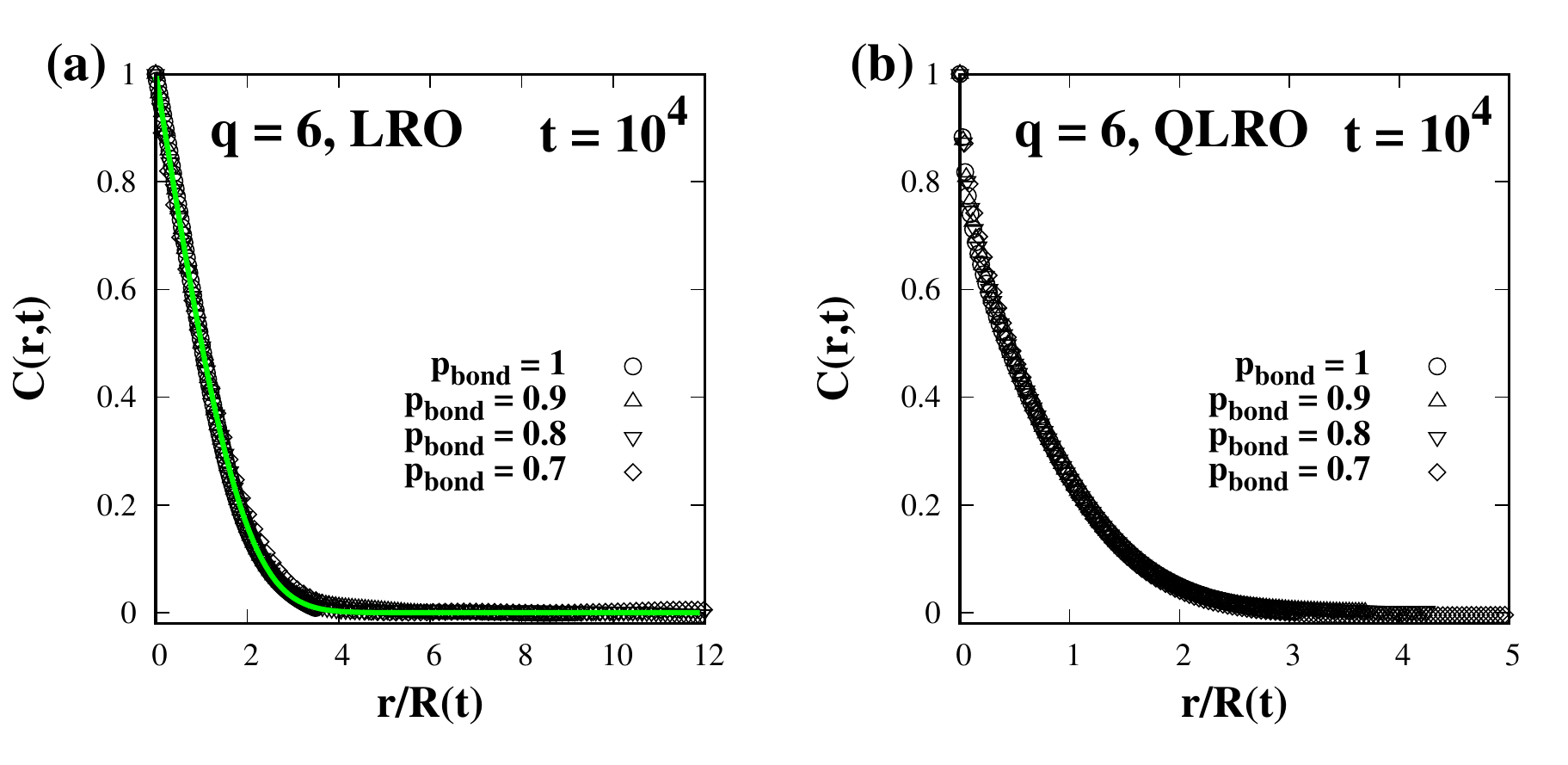}
\caption{Dynamical scaling of the correlation function in BDCM at $t = 10^4$. (a) scaled $C(\textbf{r},t)$ versus $r/R(t)$ for a quench in the LRO regime, and (b) scaled correlation function for a quench in the QLRO regime. The green solid master (color online) curve in (a) signify the Bray-Puri-Toyoki (BPT) function.}
\label{fig10}
\end{figure}

Fig.~\ref{fig10}(a) and Fig.~\ref{fig10}(b) show the dynamical scaling of the spatial correlation functions in the LRO and QLRO regimes respectively obtained at $t=10^4$ for different $p_{bond}=1, 0.9, 0.8, 0.7$. The scaling in Fig.~\ref{fig10}(a) and Fig.~\ref{fig10}(b) are analogous to the RBCM demonstrated in Fig.~\ref{fig6}(a) and Fig.~\ref{fig6}(c) but for different quenched disorder. In accordance with the RBCM, we notice that SU is also valid in BDCM and the BPT function fits nicely with the data shown in Fig.~\ref{fig10}(a). However, the BPT function does not fit well with Fig.~\ref{fig10}(b) which is a signature of the non-Porod behavior of the scaled correlation functions discussed in Sec.~\ref{ss5}.

\begin{figure}[htbp]
\centering
\includegraphics[width=\columnwidth]{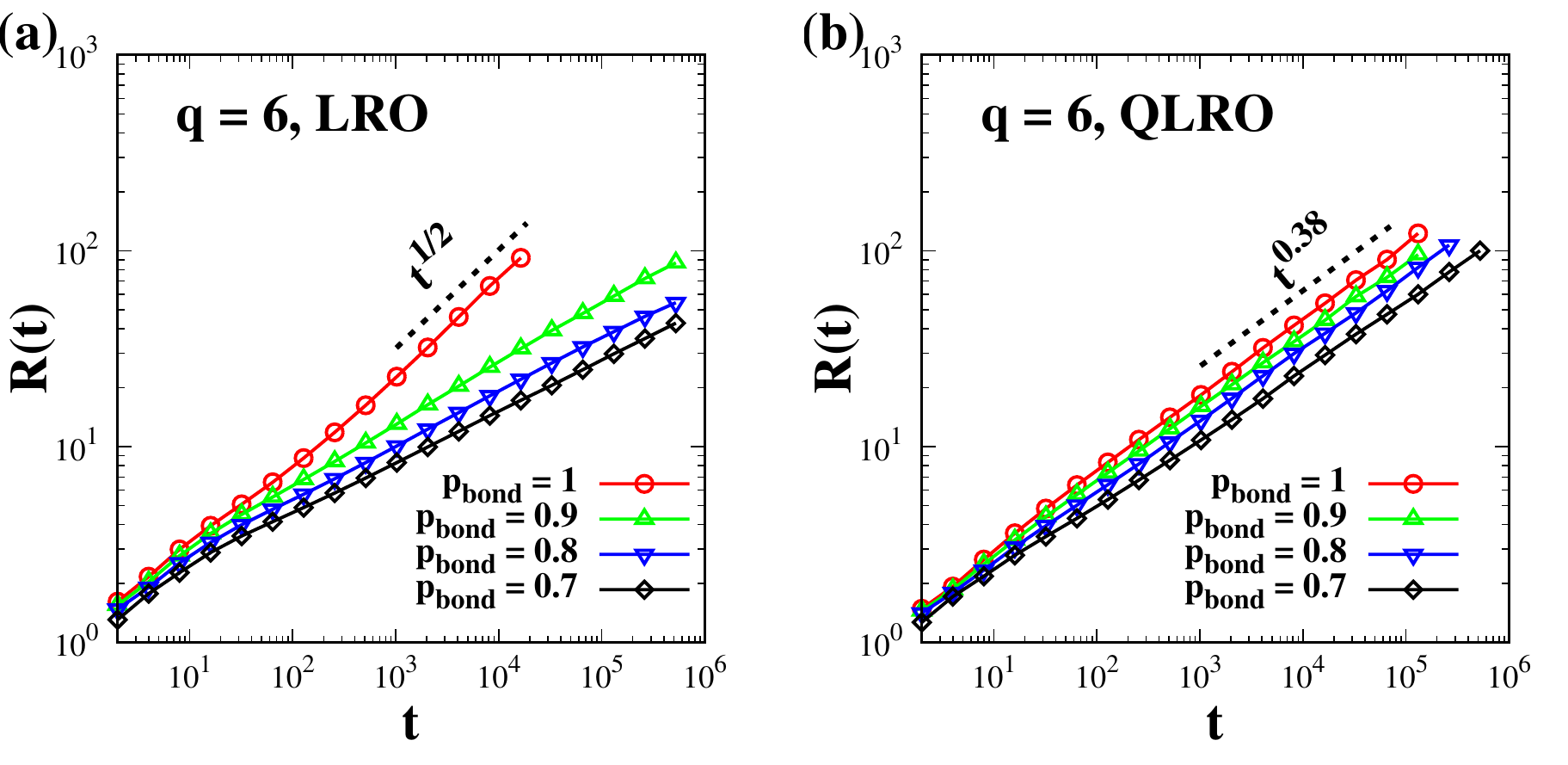}
\caption{(Color online) Length scale $R(t)$ versus $t$ (on a log-log scale) for $q$ = 9, quench temperature $T=0.1$ and $p_{bond}$ = 1 (red open circle), 0.9 (green open triangle), 0.8 (blue open inverted triangle), and 0.7 (black open rhombus). (a) Domain growth for quench in the LRO regime. Dotted line indicates the growth law of the pure clock model $R(t) \sim t^{1/2}$. (b) Domain growth for quench in the QLRO regime where the dotted line indicates growth exponent of the pure clock model $\psi_{QLRO}(p_{bond}=1) \sim 0.38$. }
\label{fig11}
\end{figure}

The average length scale $R(t)$ for BDCM, shown in Fig.~\ref{fig11} for $p_{bond}=1$ (red open circle),0.9 (green open triangle), 0.8 (blue open inverted triangle), 0.7 (black open rhombus), are calculated from the decay of the correlation function to 0.2 of its maximum value. The LRO length scale data shown in Fig.~\ref{fig11}(a) shows systematic decrease in the domain size with disorder where in the pure case ($p_{bond}=1$), guided by the dotted lines, validates the LCA growth law $R(t) \sim t^{1/2}$. The growth exponents corresponding to the subsequent $p_{bond}$ are tabulated in Table~\ref{table3}. In the QLRO regime, shown in Fig.~\ref{fig11}(b), the pure case growth exponent $\psi_{QLRO}(p_{bond}=1) \sim 0.38$ is much less than the LRO regime is consistent with the previous findings \cite{swarna2018,corberi2006} (see the Appendix). These exponents suggest that the effect of bond-dilution on the coarsening dynamics of clock model is stronger when the system is quenched to the LRO regimes compared to the QLRO regimes and this finding is consistent with our observation of the RBCM scenario presented in Sec.~\ref{ss5}.

\begin{table}[htbp]
\centering
\caption{Growth exponents $\psi_{LRO}(p_{bond})$ and $\psi_{QLRO}(p_{bond})$ for $q=6$ BDCM.}
\vspace{0.1in} 
\label{table3}
\scalebox{1.1}{
\begin{tabular}{ | p{1.0 cm} | c | c |}
\hline
$p_{bond}$ & $\psi_{LRO}(p_{bond})$ & $\psi_{QLRO}(p_{bond})$ \\
\hline
1 & 0.503 $\pm$ 0.006 & 0.389 $\pm$ 0.004 \\
\hline
0.9 & 0.296 $\pm$ 0.003 & 0.367 $\pm$ 0.002 \\
\hline
0.8 & 0.264 $\pm$ 0.003 & 0.365 $\pm$ 0.002 \\
\hline
0.7 & 0.261 $\pm$ 0.001 & 0.356 $\pm$ 0.001\\
\hline
\end{tabular}
}
\end{table}
\section{Summary and Discussion}
\label{summary}
For the past few decades domain growth in disordered systems have been a great subject of interest to statistical physicists and examples include domain growth in disordered magnets \cite{manoj2017,puri91,puri93,rp2004,rp2005,henkel2006,rp2007,henkel2008,puri2010,puri2011,puri2012,
manoj2014,manoj17,martin84,bray-humayun,rieger96,biswal96,aron2008,cugliandolo2010,park2010,pleimling2012,arun2016,hartmann2017,
manojpotts2018}, disordered type-II superconductors \cite{nicodemi2002,olson2003,gregory2004,bustingorry2006,bustingorry2007,du2007,pleimling2011}, or polymers in random media \cite{kolton2005,noh2009,cugliandolo2009,monthus2009}. We can now safely claim that we have a reasonable understanding of the physics of ordering kinetics in disordered media although authentic theoretical equipment or experimental studies have not kept pace with the numerical developments. In addition to the lack of theoretical or experimental support, coarsening dynamics in disordered systems sometimes become extremely slow and characteristic length scale becomes very small within the numerically accessible time window, as seen in spin glasses \cite{vincent,rieger2004}. The most crucial quantity one investigates in an ordering kinetics problem is the growing length scale $R(t)$ which is a function of disorder and debates are there whether it grows logarithmically \cite{HH} with time or sustains an algebraic growth \cite{manoj2017,rp2004,rp2005}. However, recent numerical developments acknowledge that the algebraic growth is transient and there happens a late time crossover from a pre-asymptotic faster algebraic growth regime to an asymptotic slower logarithmic regime in presence of disorder \cite{puri2011,puri2012,pleimling2012}.

In this work, we have undertaken a comprehensive Monte Carlo simulations of domain growth in $q$-state clock model with the quenched bond disorder (RBCM) and non-conserved (Glauber) spin-flip kinetics. In this model, the nearest neighbor coupling between clock spins $\{J_{ij}\}$, are chosen from a uniform distribution [$1-\epsilon/2$, $1+\epsilon/2$], with $\epsilon$ measuring the amplitude of disorder. $\epsilon=0$ retrieves the pure clock model and $\epsilon=2$ signifies the maximum bond-disorder for ferromagnetic interaction. An interesting fact about the $q$-state clock model for $q \geqslant 5$ is the dual-phase transitions occurring from disordered to QLRO phase at $T_c^1$ and from QLRO to LRO phase at $T_c^2$ \cite{kadanoff77,plascak2010}. We first investigate the equilibrium picture of the RBCM for $q$ = 6 and 9 and quantify $T_c^1$ and $T_c^2$ as a function of $\epsilon$. Our data suggest a systematic decrease in the transition temperatures as $\epsilon$ increases, akin to the observations made earlier in RBIM \cite{rp2005} and RBXYM \cite{manoj2017}. $T_{c}^1(\epsilon)$ are characterized from the Binder cumulant $U_4$ versus $T$ and $T_{c}^2(\epsilon)$ are extracted from the temperature dependence of $U_m$, defined in the same spirit of $U_4$ in Eq.~\ref{um}. $T_c^1(\epsilon)$ and $T_c^2(\epsilon)$ for $q$ = 6 and 9 are tabulated in Table \ref{table1}. This investigation enables us to locate the temperature quench regimes required to study the coarsening dynamics in the clock model under the influence of $\epsilon$.

Domain growth kinetics in RBCM is studied by preparing the system at temperature $T=\infty$ and then independently quenching at temperatures (a) $T<T_c^2(\epsilon)$ (LRO regime) and (b) $T_c^2(\epsilon)<T<T_c^1(\epsilon)$ (QLRO regime). Domain morphologies for various $\epsilon$ are characterized qualitatively and quantitatively by equal time spatial correlation function $C(\vec{r},t)$ and its Fourier transform, the structure factor $S(k,t)$. A quench in the LRO regime is marked by well defined, sharp domain interfaces where domain size decreases with the disorder. A similar picture, however, is not obvious when the quench is made in the QLRO regime - interpenetrating, non-compact domains with rough domain interfaces is the primary characteristic in this regime. We verify that in resonance with the RBIM and RBXYM, RBCM data also supports dynamical scaling in terms of correlation function $C(\vec{r},t)$ and structure factor $S(k,t)$, both of which are time and disorder invariant. For a quench in the LRO regime, the large-$k$ behavior of the structure factor tail falls in between the Porod decay ($n=1$) and generalized Porod law ($d=n=2$), whereas quench in the QLRO regime is defined by the non-Porod behavior of the structure factor tail. Our analysis of the length scale data, for quench in the LRO and QLRO regimes, yields a power-law growth with temperature and disorder dependent growth exponents within the simulation time scales. This feature is similar to the intermediate-time behavior for ordering kinetics in RBIM \cite{puri2010,puri2011} and asymptotic behavior in ordering kinetics in $d=2$ RBXYM \cite{manoj2017}. The quench in the LRO regime is further characterized by a power-law fit of the effective exponent $\bar{z}$ $[\sim$ $1/\psi_{LRO}(\epsilon)]$ with $\epsilon$, whereas for a quench in the QLRO regime, the fit $\bar{z}^\prime$ $[\sim 1/\psi_{QLRO}(\epsilon)$] is linear with $\epsilon$.

To present a broad picture of the domain growth of $q$-state clock model influenced by the quenched disorder, we have also explored the coarsening dynamics in the bond-diluted clock model where bonds are withdrawn from the square lattice in a probabilistic manner. Our findings in the BDCM are in resonance with the findings of the RBCM. Sharp domain boundaries for a quench in the LRO regime and rough interpenetrating domains for a quench in the QLRO regime are the salient features of the domain morphologies. Once again, we find the dynamical scaling to be super-universal and observe that the effect of disorder on the length scale is more significant in the LRO regime compared to the QLRO regime and the growth exponents are disorder-dependent.

Our present investigation of RBCM along with previous results \cite{swarna2018} provides a comprehensive understanding of the ordering kinetics in $q$-state clock model with/without the disorder. Now, the rich physics of the dynamical version of the discrete clock model or its continuum version, the XY model, has been used to study the collective motion or flocking behavior in several systems \cite{vicsek,tt,aim}. The dynamical XY model surprisingly shows an LRO phase in the low-temperature regime and due to its continuous rotational symmetry, could explain the coherent collective motion of a group of birds \cite{tt}. The dynamical version of the $q$-state clock model with $q=2$ (Active Ising Model or AIM) also proved very useful in explaining the liquid-gas phase transition with an intermediate co-existence phase \cite{aim}. Effect of quenched disorder (random-field or random-bond) on the ordering dynamics of self-propelled particles has also gained significant interest in recent times \cite{rfam}. Therefore, we hope, that in future the active version of the clock model (both presence and absence of quenched disorder) with $q>2$ would be useful to probe rich phase transitions in the field of active matter physics. 

Apart from this, an interesting problem that could arise in the context of domain growth in the clock model and XY model is due to the annealed disorder. In the present study of RBIM, we have only considered a quenched disorder, where, the impurities remain fixed at disorder sites and do not equilibrate with the host. However, an annealed disorder allows the host and the impurities to remain in thermal equilibrium as the nonrigid impurities are not fixed in time \cite{LF}. Fundamentally, the relaxation time associated with the diffusion of the impurities is much larger in the quenched case compared to the annealed scenario. Apart from a few studies \cite{sanjay,parongama}, a more common practice has been the study of domain kinetics with the quenched disorder. We believe that coarsening dynamics in the clock model or the XY model with the annealed disorder would be interesting to investigate in the future.

In a similar context, Kibble-Zurek (KZ) mechanism \cite{KZ}, which is very well known both to the cosmology and condensed matter communities, can also be exploited to study the coarsening dynamics. This mechanism is an equilibrium scaling argument which estimates the density of topological defects as a function of the finite rate cooling after the quench. Although the majority of theoretical studies involve rapid quench below the transition temperature, in experiments, such quenches are generally performed at a finite rate and therefore the KZ mechanism is very relevant. Clock model and XY model involve topological defects such as vortices and anti-vortices, and in these systems, ordering kinetics is driven via the annihilation of such defects $i.e.$ defect density is a decreasing function of time and domain sizes. KZ mechanism examined in $d=2$ pure XY model for vortex density by cooling through the Kosterlitz-Thouless transition point at a finite rate, suggests that the quench rate dependence in systems like the XY model goes beyond the equilibrium scaling arguments \cite{leticia}. Therefore, as a future course of the investigation, it would be interesting to investigate the KZ mechanism under slow annealing in the clock and XY model with the disorder.

\begin{acknowledgments} 
S.C. thanks CSIR, India, for support through Grant No. 09/080(0897)/2013-EMR-I and Indian Association for the Cultivation of Science, Kolkata for financial support. R.P. thanks CSIR, India, for support through Grant No. 03(1414)/17/EMR-II. S.P. is grateful to the Department of Science and Technology, India, for funding through a J.C. Bose fellowship.
\end{acknowledgments}

\appendix*

\section{Coarsening dynamics in RBCM with $q$ = 4 and 6}
\label{appndx}

In Sec.~\ref{ss5}, we have shown that ordering kinetics in $q=9$ state RBCM can be best described by power-law domain growth with disorder-dependent growth exponents for quenches in the LRO and QLRO phases. In this section, we show that the outcome is consistent with the coarsening for other $q$ values.

\subsection{Quench in the LRO phase}
\begin{figure}[htbp]
\centering
\includegraphics[width=\columnwidth]{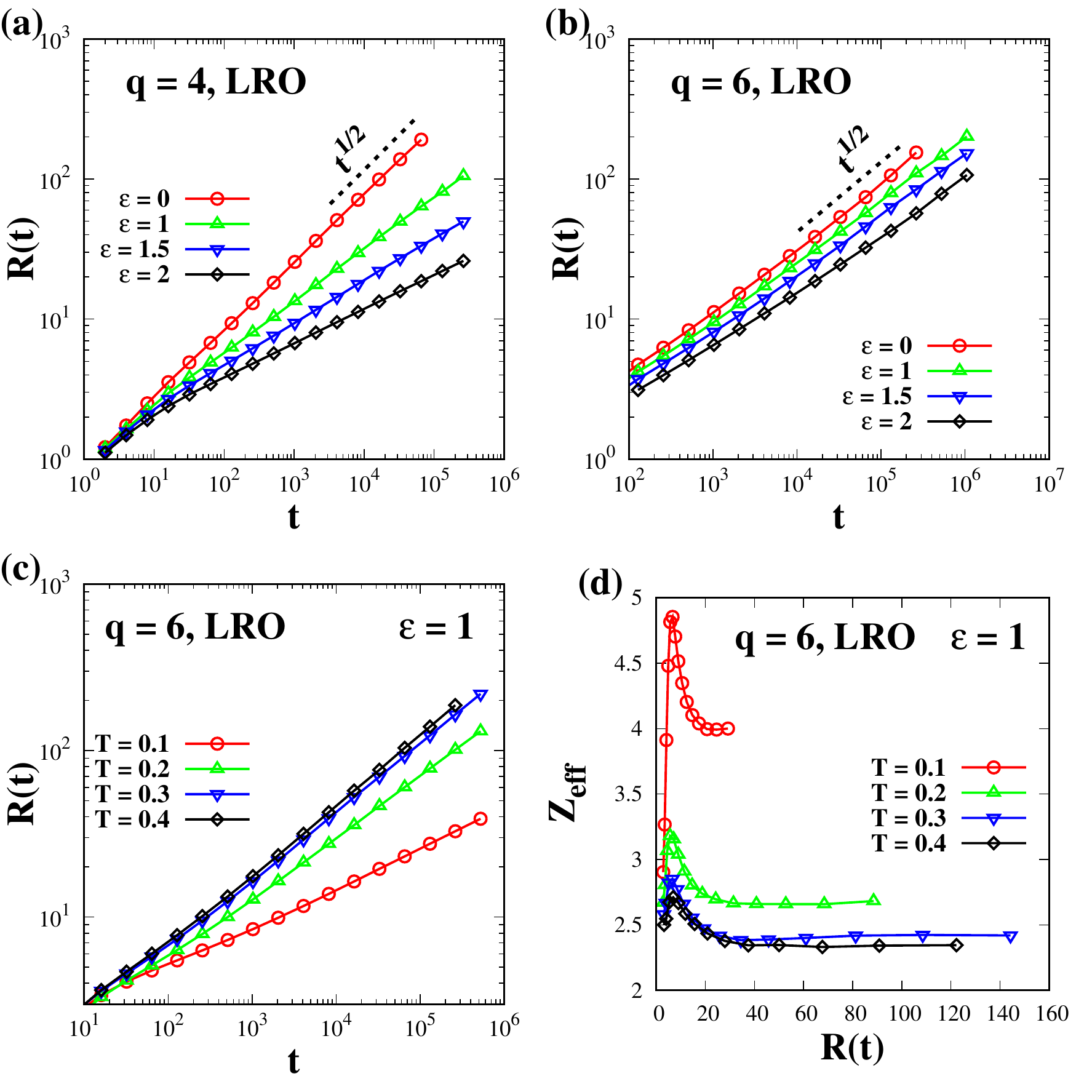}
\caption{(Color online) Length scale data for a quench in the LRO regime. (a) Plot of $R(t)$ versus $t$ (on a log-log scale) for $q=4$ and $\epsilon$ = 0 (red open circle), 1 (green open triangle), 1.5 (blue open inverted triangle) and 2 (black open rhombus) after a quench from $T = \infty$ to $T = 0.4$. The dashed line indicates the growth law $R(t) \sim t^{1/2}$ and is provided as a guide to the eye. (b) Analogous to (a) but for $q=6$ and a quench to $T=0.5$. (c) Plot of $R(t)$ versus $t$ (on a log-log scale) for $q=6$ and $\epsilon=1$ with different quench temperatures $T$ = 0.1 (red open circle), 0.2 (green open triangle), 0.3 (blue open inverted triangle), and 0.4 (black open rhombus) in the LRO regime. (d) Effective exponent $Z_{eff}$ versus $R(t)$ plot corresponding to the data shown in (c).}
\label{figA1}
\end{figure}

In $q$-state clock model, QLRO phase begins with $q \geqslant5$ and therefore no QLRO phase associated with $q=4$ \cite{kadanoff77,plascak2010}. We first find that $T_c$ for $q=4$ decreases from $T_c (\epsilon=0)= 1.133 \pm 0.001$ to $T_c(\epsilon=2)=1.008 \pm 0.005$. $T_c$ for $\epsilon=0$ is consistent with the prediction that $T_c (q=4)=\frac{1}{2} T_c(q=2)$ \cite{plascak2010}. Fig.~\ref{figA1}(a) shows the length scale data $R(t)$ versus $t$ for $q=4$ on a log-log scale for various $\epsilon$ after a quench in the LRO phase with quench temperature $T=0.4$. The data shows that for $q=4$ and $\epsilon=0$, the domain size $R(t) \sim t^{1/2}$; however, the growth eventually slows down at higher $\epsilon$ as reported earlier for $q=9$. $R(t)$ versus $t$ (on a log-log scale) for $q=6$ is shown in Fig.~\ref{figA1}(b) for a quench at temperature $T=0.5 (T<T_c^2)$. The plot shows that growth is disorder dependent and has been affected significantly by the disorder amplitude $\epsilon$. In Fig.~\ref{figA1}(c), we demonstrate the effect of quench temperature $T$ on the coarsening dynamics of $q=6$ state clock model for $\epsilon=1$. Domain size increases with temperature as indicated by the data at $T=0.4$ having larger domains compared other quench temperatures. Higher quench temperature signifies more thermal fluctuations which help the trapped domains to overcome the energy barriers at non-zero $\epsilon$. Studying Fig.~\ref{figA1}(a)-(c) we find that growth exponents in RBCM are both temperature- and disorder-dependent. Our data in Fig.~\ref{figA1}(d) shows stable exponents (flat regimes) corresponding to various $T$ and supports a power-law behavior of the domain coarsening where $Z_{eff}$ is defined in Eq.~\eqref{zeff}.

\subsection{Quench in the QLRO phase}
\begin{figure}[htbp]
\centering
\includegraphics[width=\columnwidth]{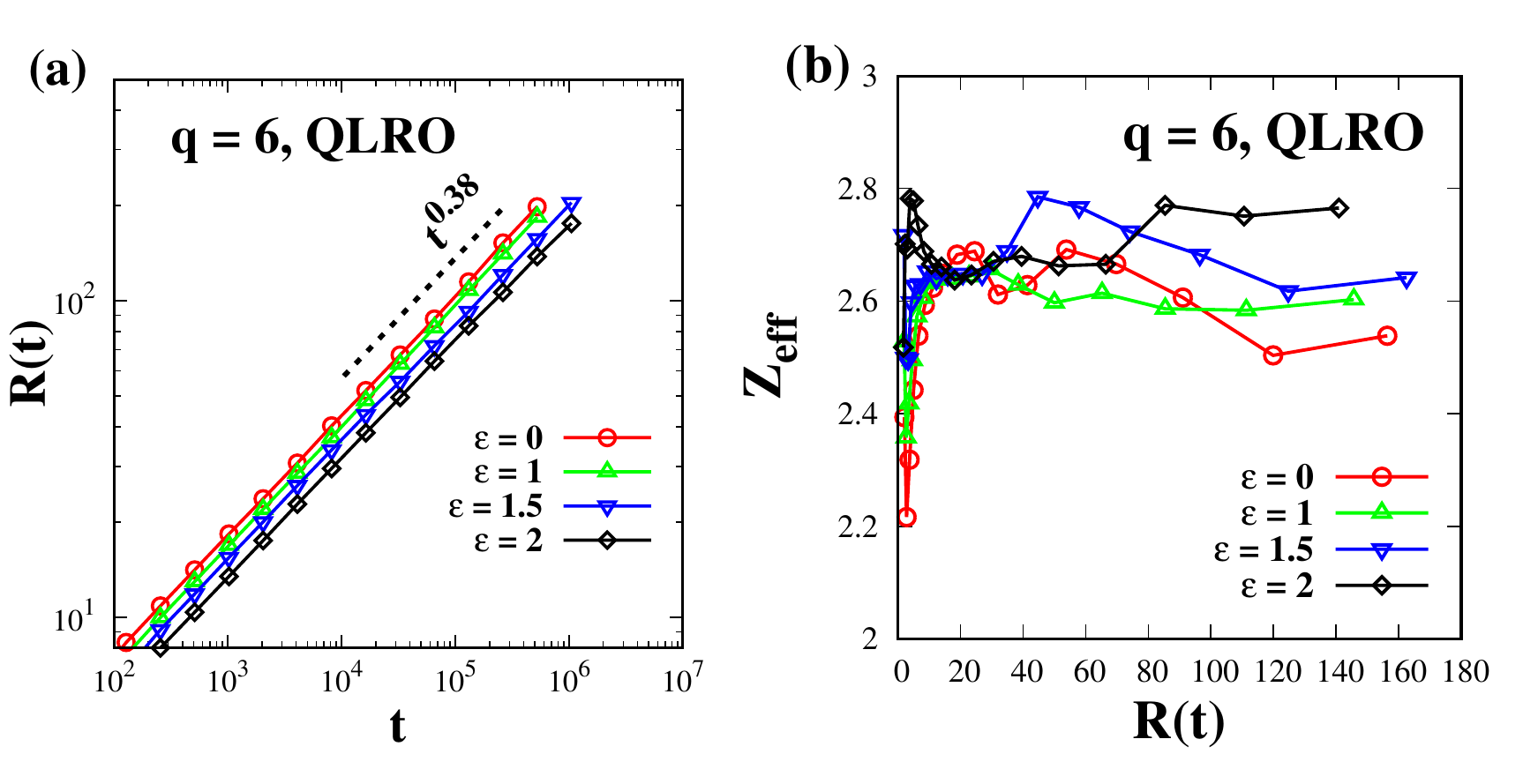}
\caption{(a)(Color online) $R(t)$ versus $t$ (on a log-log scale) for 6-state clock model after a quench from $T = \infty$ to the QLRO regime (quench temperature $T \simeq 0.9T_c^1(\epsilon)$), for specified values of $\epsilon$, $\epsilon$ = 0 (red open circle), 1 (green open triangle), 1.5 (blue open inverted triangle) and 2 (black open rhombus). The dashed line indicate the pure ($\epsilon=0$) growth law $R(t)\sim t^{0.38}$. (b) $Z_{eff}$ versus $R(t)$ corresponding to the data presented in (a).}
\label{figA2}
\end{figure}

In order to study the coarsening for $q=6$ in the QLRO phase, the system is quenched at temperature $T=0.9T_c^1(\epsilon)$, where $T_c^1(\epsilon)$ are tabulated in Table~\ref{table1}. The characteristic length scale $R(t)$ versus $t$ (on a log-log scale) for different $\epsilon$ is plotted in Fig.~\ref{figA2}(a). The pure case ($\epsilon=0$) shows an algebraic domain growth with exponent $\sim$ 0.38 \cite{swarna2018}. Subsequent data for higher $\epsilon$ suggests that, although domain sizes corresponding to a fixed $t$ decreases with $\epsilon$, the effect of $\epsilon$ on the domain growth exponent is nominal which can also be quantified by fitting an appropriate function with the data $[\psi ({\epsilon=0}) \simeq 0.386 \pm 0.003$ to $\psi ({\epsilon=2}) \simeq 0.363 \pm 0.002]$. The $Z_{eff}$ versus $R(t)$ plot in Fig.~\ref{figA2}(b) corresponding to the data in (a) also reflects the disorder dependence of the $q=6$ RBCM quenched in the QLRO phase; however, due to lack of statistics it is difficult to distinguish the difference in growth exponents as a function of $\epsilon$.

\clearpage


\begin{thebibliography}{}
\bibitem{berezinskii70} 
V. L. Berezinskii, Sov. Phys. JETP 32, 493 (1971); Sov. Phys. JETP 34, 610 (1972). 

\bibitem{thouless73} 
J. M. Kosterlitz and D. J. Thouless, J. Phys. C 6, 1181 (1973). 

\bibitem{kosterlitz74} 
J. M. Kosterlitz, J. Phys. C 7, 1046 (1974).

\bibitem{mermin} 
N. D. Mermin and H. Wagner, Phys. Rev. Lett. 17, 1133 (1966).

\bibitem{kadanoff77} 
J. V. Jos{\'e}, L. P. Kadanoff, S. Kirkpatrick, and D. R. Nelson, Phys. Rev. B 16, 1217 (1977).

\bibitem{elitzur} 
S. Elitzur, R. B. Pearson, and J. Shigemitsu, Phys. Rev. D 19, 3698 (1979).

\bibitem{domany} 
E. Domany, D. Mukamel, and A. Schwimmner, J. Phys. A 13, L311 (1980).

\bibitem{cardy} 
J. L. Cardy, J. Phys. A 13, 1507 (1980).

\bibitem{tobochnik82} 
J. Tobochnik, Phys. Rev. B 26, 6201 (1982).

\bibitem{plascak2010} 
A. F. Brito, J. A. Redinz, and J. A. Plascak, Phys. Rev. E 81, 031130 (2010).

\bibitem{baek2010} 
S. K. Baek and P. Minnhagen, Phys. Rev. E 82, 031102 (2010).

\bibitem{swarna2018} 
S. Chatterjee, S. Puri, and R. Paul, Phys. Rev. E 98, 032109 (2018).

\bibitem{surungan2005} 
T. Surungan and Y. Okabe, Phys. Rev. B 71, 184438 (2005).

\bibitem{tomita2002} 
Y. Tomita and Y. Okabe, Phys. Rev. B 66, 180401(R) (2002).

\bibitem{miyashita78}
S. Miyashita, H. Nishimori, A. Kuroda, and M. Suzuki, Prog. Theor. Phys. 60, 1669 (1978).

\bibitem{landau86}
M. S. S. Challa and D. P. Landau, Phys. Rev. B 33, 437 (1986).

\bibitem{ono91}
A. Yamagata and I. Ono, J. Phys. A 24, 265 (1991).

\bibitem{okabe2002}
Y. Tomita and Y. Okabe, Phys. Rev. B 65, 184405 (2002).

\bibitem{tomita2001}
Y. Tomita and Y. Okabe, Phys. Rev. Lett. 86, 572 (2001).

\bibitem{rastelli2004}
E. Rastelli, S. Regina, and A. Tassi, Phys. Rev. B 69, 174407 (2004).

\bibitem{kim2010}
S. K. Baek, P. Minnhagen, and B. J. Kim, Phys. Rev. E 81, 063101 (2010).

\bibitem{Wu2012}
Raymond P. H. Wu , Veng-cheong Lo, H. Huang, J. Appl. Phys. 112, 063924 (2012).

\bibitem{li}
Zi-Qian Li, Li-Ping Yang, Z. Y. Xie, Hong-Hao Tu, Hai-Jun Liao, T. Xiang, arXiv:1912.11416 [cond-mat.stat-mech] (2020).

\bibitem{leonel2003} 
S. A. Leonel, P. Z. Coura, A. R. Pereira, L. A. S. M{\'o}l, and B. V. Costa, Phys. Rev. B 67, 104426 (2003).

\bibitem{alonso2010} 
J. J. Alonso, J. Magn. Magn. Mater. 322, 1330 (2010).

\bibitem{harris74} 
A. B. Harris, J. Phys. C 7, 1671 (1974).

\bibitem{wysin2005} 
G. M. Wysin, A. R. Pereira, I. A. Marques, S. A. Leonel, and P. Z. Coura, Phys. Rev. B 72, 094418 (2005).

\bibitem{manoj2017} 
M. Kumar, S. Chatterjee, R. Paul, and S. Puri, Phys. Rev. E 96, 042127 (2017).

\bibitem{Bray94} 
A. J. Bray, Adv. Phys. 43, 357 (1994).

\bibitem{puri-wadhawan} 
S. Puri, \textit{Kinetics of Phase Transitions}, edited by S. Puri and V. K. Wadhawan (Taylor \& Francis, Boca Raton, FL, 2009).

\bibitem{ls}
I. M. Lifshitz and V. V. Slyozov, J. Phys. Chem. Solids 19, 35 (1961).

\bibitem{lca}
I. M. Lifshitz, Sov. Phys. JETP 15, 939 (1962); S. E. Allen and J. W. Cahn, Acta. Metall. 27, 1085 (1979).

\bibitem{oono-puri}
Y. Oono and S. Puri, Phys. Rev. Lett. 58, 863 (1987).

\bibitem{BR94} 
A. J. Bray and A. D. Rutenberg, Phys. Rev. E 49, R27 (1994).

\bibitem{RB95} 
A. D. Rutenberg and A. J. Bray, Phys. Rev. E 51, 5499 (1995).

\bibitem{puri95} 
S. Puri, A. J. Bray, and F. Rojas, Phys. Rev. E 52, 4699 (1995).

\bibitem{corberi2006} 
F. Corberi, E. Lippiello, and M. Zannetti, Phys. Rev. E 74, 041106 (2006).

\bibitem{puri91} 
S. Puri, D. Chowdhury, and N. Parekh, J. Phys. A 24, L1087 (1991).

\bibitem{puri93} 
S. Puri and N. Parekh, J. Phys. A: Math. Gen. 25, 4127 (1992); J. Phys. A: Math. Gen. 26, 2777 (1993).

\bibitem{rp2004} 
R. Paul, S. Puri, and H. Rieger, Europhys. Lett. 68, 881 (2004).

\bibitem{rp2005} 
R. Paul, S. Puri, and H. Rieger, Phys. Rev. E 71, 061109 (2005).

\bibitem{henkel2006} 
M. Henkel and M. Pleimling, Europhys. Lett. 76, 561 (2006).

\bibitem{rp2007}
R. Paul, G. Schehr, H. Rieger, Phys. Rev. E75, 030104(R)(2007).

\bibitem{henkel2008} 
M. Henkel and M. Pleimling, Phys. Rev. B 78, 224419 (2008).

\bibitem{puri2010} 
E. Lippiello, A.Mukherjee, S. Puri, and M. Zannetti, Europhys. Lett. 90, 46006 (2010).

\bibitem{puri2011} 
F. Corberi, E. Lippiello, A.Mukherjee, S. Puri, and M. Zannetti, J. Stat. Mech. (2011) P03016.

\bibitem{puri2012} 
F. Corberi, E. Lippiello, A. Mukherjee, S. Puri, and M. Zannetti, Phys. Rev. E 85, 021141 (2012).

\bibitem{manoj2014} 
G. P. Shrivastav, M. Kumar, V. Banerjee, S. Puri, Physical Review E 90, 032140 (2014).

\bibitem{manoj17} 
M. Kumar, V. Banerjee, S. Puri, Euro. Phys. Lett. 117, 10012 (2017).

\bibitem{reichhardt}
C. Reichhardt and C. J. Olson Reichhardt, Phys. Rev. E 73, 046122 (2006).

\bibitem{lupo}
C. Lupo and F. Ricci-Tersenghi, Phys. Rev. B 95, 054433 (2017).

\bibitem{wolff89} 
U. Wolff, Phys. Rev. Lett. 62, 361 (1989).

\bibitem{metropolis53} 
N. Metropolis, A. W. Rosenbluth, M. N. Rosenbluth, A. H. Teller and E. Teller, J. Chem Phys. 21, 1087 (1953).

\bibitem{binder2005} 
D. P. Landau and K. Binder, \textit{A Guide to Monte Carlo Simulations in Statistical Physics} (Cambridge University Press, Cambridge, 2009).

\bibitem{baek2009} 
S. K. Baek, P. Minnhagen and B. J. Kim, Phys. Rev. E 80, 060101(R) (2009).

\bibitem{glauber} 
R. J. Glauber, J. Math. Phys. \textbf{4}, 294 (1963).

\bibitem{kawasaki}
K. Kawasaki, in Phase Transition and Critical Phenomena, edited by C. Domb and M. S. Green (Academic, New York, 1972), Vol. 2, p. 443.

\bibitem{binder-stauffer} 
K. Binder and D. Stauffer, Phys. Rev. Lett., 33, 1006 (1974).

\bibitem{BP} 
A. J. Bray and S. Puri, Phys. Rev. Lett., 67, 2670 (1991).

\bibitem{T} 
H. Toyoki, Phys. Rev. B, 45, 1965 (1992).

\bibitem{bray-humayun}
A. J. Bray and K. Humayun, J. Phys. A: Math. Gen. 24, L1185 (1991).

\bibitem{sicilia}
A. Sicilia, J. J. Arenzon, A. J. Bray, and L. F. Cugliandolo, Europhys. Lett. 82, 10001 (2008).

\bibitem{subir}
S. Ahmad, S. Puri, and S. K. Das, Phys. Rev. E 90, 040302(R) (2014).

\bibitem{puri2016} 
A. Bupathy, V. Banerjee, and S. Puri, Phys. Rev. E 93, 012104 (2016).

\bibitem{puri97} 
S. Puri, R. Ahluwalia, and A. J. Bray, Phys. Rev. E 55, 2345 (1997).

\bibitem{HH} 
D. A. Huse and C. L. Henley, Phys. Rev. Lett. 54, 2708 (1985).

\bibitem{martin84}
M. Grant and J. D. Gunton, Phys. Rev. B 29, 1521(R) (1984).

\bibitem{rieger96}
H. Rieger, Physica A 224, 267 (1996).

\bibitem{biswal96}
B. Biswal, S. Puri, D. Chowdhury, Physica A 229, 72 (1996).

\bibitem{aron2008}
C. Aron, C. Chamon, L.F. Cugliandolo, M. Picco, J. Stat.Mech. P05016 (2008).

\bibitem{cugliandolo2010}
M. P. O. Loureiro, J. J. Arenzon, L. F. Cugliandolo, A. Sicilia, Phys. Rev. E 81, 021129 (2010).

\bibitem{park2010}
H. Park, M. Pleimling, Phys. Rev. B82, 144406 (2010).

\bibitem{pleimling2012}
H. Park and M. Pleimling, Eur. Phys. J. B 85, 300 (2012).

\bibitem{arun2016}
A. Bupathy, V. Banerjee, and S. Puri, Phys. Rev. E 93, 012104 (2016).

\bibitem{hartmann2017}
S. von Ohr, M. Manssen, and A. K. Hartmann, Phys. Rev. E 96, 013315 (2017).

\bibitem{manojpotts2018}
M. Kumar, R. Kumar, M. Weigel, V. Banerjee, W. Janke, and S. Puri, Phys. Rev. E 97, 053307 (2018).

\bibitem{nicodemi2002}
M. Nicodemi, H. J. Jensen, Phys. Rev. B 65, 144517 (2002).

\bibitem{olson2003}
C. J. Olson, C. Reichhardt, R. T. Scalettar, G. T. Zimanyi, N. Gr{\o}nbach-Jensen, Phys. Rev. B 67, 184523 (2003).

\bibitem{gregory2004}
G. Schehr, P. Le Doussal, Phys. Rev. Lett. 93, 217201 (2004).

\bibitem{bustingorry2006}
S. Bustingorry, L. F. Cugliandolo, D. Dominguez, Phys. Rev. Lett. 96, 027001 (2006).

\bibitem{bustingorry2007}
S. Bustingorry, L. F. Cugliandolo, D. Dominguez, Phys. Rev. B 75, 024506 (2007).

\bibitem{du2007}
X. Du, G. Li, E.Y. Andrei, M. Greenblatt, P. Shuk, Nature Physics 3, 111 (2007).

\bibitem{pleimling2011}
M. Pleimling, U. C. T{\"a}uber, Phys. Rev. B 84, 174509(2011).

\bibitem{kolton2005}
A. Kolton, A. Rosso, T. Giamarchi, Phys. Rev. Lett. 95, 180604 (2005). 

\bibitem{noh2009} 
J. D. Noh, H. Park, Phys. Rev. E 80, 040102(R) (2009).

\bibitem{cugliandolo2009} 
J. L. Iguain, S. Bustingorry, A. B. Kolton, L. F. Cugliandolo, Phys. Rev. B 80, 094201 (2009). 

\bibitem{monthus2009} 
C. Monthus, T. Garel, J. Stat. Mech. P12017 (2009).

\bibitem{vincent}
E. Vincent, in \textit{Ageing and the Glass Transition}, edited by M. Henkel, M. Pleimling, R. Sanctuary (Springer, Heidelberg, 2007). 

\bibitem{rieger2004}
N. Kawashima and H. Rieger, in \textit{Frustrated Magnetic Systems}, edited by H. Diep (World Scientific, Singapur, 2004).

\bibitem{vicsek} 
T. Vicsek, A. Czir{\'o}k, E. Ben-Jacob, I. Cohen, O. Shochet, Phys. Rev. Lett. 75, 1226 (1995).

\bibitem{tt} 
J. Toner and Y. Tu, Phys. Rev. Lett. 75, 4326 (1995); Phys. Rev. E 58, 4828 (1998).

\bibitem{aim} 
A. P. Solon and J. Tailleur, Phys. Rev. Lett. 111, 078101 (2013); Phys. Rev. E 92, 042119 (2015).

\bibitem{rfam}
R. Das, S. Mishra, and S. Puri, Europhys. Lett. 121, 37002 (2018). 

\bibitem{LF}
L. F. Cugliandolo, Disordered Systems, Lecture notes (Cargése, 2011).

\bibitem{sanjay}
S. Puri and R. Sharma, Phys. Rev. E 57, 1873 (1998).

\bibitem{parongama}
S. Biswas and P. Sen, Phys. Rev. E 80, 027101 (2009).

\bibitem{KZ}
T. W. B. Kibble, J. Phys. A 9, 1387 (1976); W. H. Zurek, Nature (London) 317, 505 (1985); W. H. Zurek, Phys. Rep. 276, 177 (1996).

\bibitem{leticia}
A. Jeli{\'c} and L. F. Cugliandolo, J. Stat. Mech. (2011) P02032.

\end{thebibliography}
\end{document}